\documentclass[pra,amsmath,amssymb,, showpacs]{revtex4}

\usepackage{graphicx}

\usepackage{epsfig}

\usepackage{float}

\begin{document}

\title{Perturbative hydrogenic Lamb shifts and radiative decay rates - an $so(4,2)$-based algebraic approach}

\author{ G.\ Alber}
\affiliation{
Institut f\"{u}r Angewandte Physik, Technische Universit\"{a}t Darmstadt, D-64289, Germany}

\date{\today}

\begin{abstract}
It is shown that algebraic techniques based on the Lie algebra $so(4,2)$ provide efficient tools for evaluating Lamb shifts and radiative decay rates for hydrogenic energy eigenstates as they systematically exploit the intrinsic symmetry of the hydrogenic Hamiltonian. As a main result in lowest order perturbation theory with respect to the fine-structure constant integral representations are derived for the complex-valued energy shifts of hydrogen-like ions from which Lamb shifts and radiative decay rates can be evaluated in a unified way, thus  generalizing a recently discussed algebraic approach of Maclay \cite{Maclay}. In order to exemplify the usefulness of this algebraic approach numerical results are presented for Lamb shifts and radiative decay rates which transcend the dipole approximation and contain the dipole approximation as a limiting case. 
\end{abstract}
\pacs{31.15.-p, 31.15.xh, 31.15.Hz, 32.70.Jz, 03.65.-w, 03.65.Fd, 03.65.Ge}

\maketitle

\section{Introduction}
Understanding the physics of the hydrogen atom has been a major impetus for the development of quantum mechanics and quantum electrodynamics. It is one of the simplest realistic quantum systems, and in view of its high symmetry  it is a paradigmatic physical system for exploring fundamental theoretical aspects \cite{Kepler}. After the early success of its non-relativistic theoretical description by Schrödinger and its relativistic generalization by Dirac in the 1920's \cite{BS} it was the experimental observation of the non-degeneracy of the $2s$ and the $2p$ levels of hydrogen in the experiment by Lamb and Retherford in 1947 \cite{Lamb} and Bethe's ground breaking non-relativistic quantum theoretical explanation \cite{Bethe} which significantly advanced the further development of quantum electrodynamics. This theoretical explanation attributed this non-degeneracy as well as the radiative decay rates of excited hydrogen states to the vacuum fluctuations of the quantized electromagnetic field.

Bethe's original calculation is based on a non-relativistic description of the hydrogen atom in the infinite mass limit. Thereby, the coupling of the electron to the transverse electromagnetic field modes is described in lowest (second) order perturbation theory in the atom-field coupling within the framework of the dipole approximation. A key novel idea  of this theoretical treatment was the subtraction of a mass renormalization term from the resulting complex-valued energy shifts which renders the remaining part, describing the Lamb shifts and radiative decay rates, only logarithmically divergent in the cut-off frequency of the virtual photons involved. Later treatments within the framework of the dipole approximation, which also take into account relativistic corrections in lowest order perturbation theory
with respect to the fine-structure constant systematically, introduced the cut-off independent Bethe-logarithm \cite{BS,Huff} for characterizing the atomic contribution to the Lamb shifts.

Early subsequent theoretical treatments of the Lamb shifts of the hydrogen atom by Lieber \cite{Lieber}
and Huff \cite{Huff}
are based on the same approximations, but in addition they also exploit the high symmetry properties of the non-relativistic hydrogen Hamiltonian algebraically. 
In Lieber's approach the $SO(4)$ symmetry of the non-relativistic hydrogenic Hamiltonian is exploited algebraically in conjunction with Schwinger's integral representation of the Coulomb Green's function \cite{Schwinger}. 
Huff \cite{Huff} extends these algebraic methods further by using the Lie algebra $so(4,2)$ and properties of its associated Lie group $SO(4,2)$. He demonstrates that by invoking this higher symmetry properties one may obtain a simpler expression for the Bethe-logarithm by taking advantage of the resulting powerful algebraic properties. This is also due to the fact that this particular representation of $so(4,2)$ also contains the electronic canonical momentum operator, which characterizes the electron's coupling to the electromagnetic field modes. By this approach Huff is able to represent the Bethe-logarithm for hydrogenic energy eigenstates with well defined angular momenta as a sum over a complete discrete set of eigenstates of one of the self-adjoint generators of $SO(4,2)$. This approach also exhibits the central role played by a particular $SO(2,1)$ subgroup of $SO(4,2)$ by demonstrating that this infinite sum just represents a particular matrix element of one of the unitary operators of this subgroup. 

A later non-relativistic calculation of hydrogen Lamb shifts by Seke and M\"odritsch \cite{Seke,Seke1} demonstrated another interesting aspect, namely that without the dipole approximation cut-off independent results are obtained for the  Lamb shifts. In particular, this approach takes into account retardation effects involved in the coupling between a hydrogenic electron and the virtual photons. These retardation effects are ignored in the dipole approximation. Thus, these results demonstrate the significance of these retardation effects even within the framework of a non-relativistic perturbative treatment of the Lamb shift.

By extending earlier group theoretical approaches based on the Lie algebra  $so(4,2)$, recently
Maclay \cite{Maclay,Maclay1} presented an interesting integral representation for the Lamb shifts of hydrogenic energy eigenstates with well defined angular momenta valid in lowest order perturbation theory within the framework of the dipole approximation. This approach is based on a particular infinite dimensional unitary representation of $so(4,2)$ and its associated Lie group $SO(4,2)$. The particular values of the Casimir operators of this representation yield a powerful contraction relation, which greatly simplifies the evaluation of the relevant matrix elements determining the
Lamb shifts. Thereby, these Lamb shifts are
represented by a double integral over the frequencies of the virtual photons involved and over an imaginary-valued effective time. The kernel of this integral representation 
involves diagonal matrix elements of an effective time evolution operator involving one of the generators of $SO(4,2)$. A particular advantage of this approach is that, given these matrix elements, the double integral can be performed in a straight forwards way numerically and no summations over contributions of intermediate states are required. A potential drawback of this approach is that these matrix elements are not defined explicitly but only implicitly as coefficients of a Taylor expansion of a generating function. 

The application of algebraic and group theoretical methods for the determination or approximate description of energy levels or energy eigenstates of atoms or molecules has a long tradition in quantum mechanics and a rich literature \cite{Wigner,Judd}. One of the first contributions in this respect is Pauli's seminal determination of the discrete spectrum of hydrogen based on the Lie algebra $so(4)$ and its associated compact Lie group $SO(4)$ \cite{Pauli}. 
An early account of the significance of the non-compact Lie group $SO(2,1)$ for describing hydrogenic energy eigenstates of well defined angular momenta has been given by Armstrong \cite{Armstrong}. Group theoretical methods based on the hydrogenic degeneracy group $SO(4)$ have also been used for the characterization of electron correlation effects in two-electron atoms where they have been applied successfully to the characterization of some discrete energy spectra \cite{Wulfmann,Herrick,Nikitin,Rau1990}.
However, algebraic methods for applications transcending approximate diagonalizations of atomic or molecular Hamiltonians, which, for example, require the determination of matrix elements of the relevant Hamiltonian resolvent, such as required for the evaluation of Lamb shifts and radiative decay rates, are not yet as well explored. 

Motivated by these recent developments and the renewed interest in the physics of the Lamb shift \cite{Maclay2,Keitel,Milonni} and in symmetry-based algebraic theoretical approaches, it is a main purpose of this paper to generalize MacLay's recent algebraic results \cite{Maclay,Maclay1} by transcending the dipole approximation and determining the relevant complex-valued energy shifts of hydrogenic energy eigenstates with well defined angular momenta, whose real parts determine their Lamb shifts and whose imaginary parts determine their radiative decay rates. Within this framework the corresponding results originating from the dipole approximation are contained as a limiting case. Thus, lowest order perturbative non-relativistic calculations of Lamb shifts and radiative decay rates without the dipole approximation can be compared in a straight forward way with the corresponding dipole-approximation based results involving the Bethe-logarithm and possibly also further relativistic quantum electrodynamic effects. Although the following approach builds on the powerful algebraic features of the particular representation of $so(4,2)$ used by Maclay \cite{Maclay,Maclay1}, even in the limit of the dipole approximation the  integral representations of the complex-valued energy shifts presented in the following differ from these earlier results in two respects. Firstly, the matrix elements involved in the kernels of the integral representations presented in the following are given explicitly in terms of particular hypergeometric functions or Jacobi polynomials. They reflect the significance of a particular $SO(2,1)$ subgroup of this representation of $SO(4,2)$. On the basis of these explicitly known integral kernels Lamb shifts and decay rates can be evaluated numerically in a straight forward way. Secondly, 
besides the required integration over the frequencies of the virtual photons
these representations involve an integration over a real-valued effective time, which may be identified with the eccentric anomaly of the Kepler problem \cite{Kepler}. By deforming the contour of this time integration appropriately in the complex time-plane these integral representations can be represented equivalently by an integration over an imaginary-valued effective time variable together with principal value and residuum contributions originating from simple poles of the integrand involved. The residuum contributions yield the radiative decay rates and the corresponding principal values contribute to the Lamb shifts. These latter principal value contributions originating from these poles are not included in the recent approach by Maclay,  but they contribute significantly to the Lamb shifts of excited hydrogenic energy eigenstates.

This paper is organized as follows. For the sake of a coherent presentation Sec. \ref{NRL} summarizes basic facts concerning lowest-order perturbative calculations of Lamb shifts and radiative decay rates of a hydrogen-like ion in the non-relativisitc approximation. Restricted to non-relativistic electron velocities this treatment transcends the dipole approximation. 
In
Sec. \ref{LieH} basic properties of the particular representation of the Lie algebra $so(4,2)$ are summarized 
which has been used and discussed in detail recently by Maclay \cite{Maclay} and
which clearly exhibits the algebraic symmetry properties of the hydrogenic Hamiltonian.
Based on this representation in 
Sec. \ref{LieR} integral representations are derived for the complex-valued energy shifts of hydrogenic eigenenergies resulting from the coupling of the hydrogenic electron to the electromagnetic radiation field in lowest order perturbation theory. These integral representations are convenient starting points for evaluating hydrogenic Lamb shifts and radiative decay rates of all energy eigenstates of definite angular momenta and are a main result of this paper. In the first subsection the general relation between the retarded resolvent of the hydrogenic Hamiltonian and the perturbative complex-valued energy shifts is established within the framework of this $so(4,2)$-based algebraic approach. 
Thereby, integral representations are derived for these complex-valued energy shifts which involve integrations over the frequencies of the relevant virtual photons and over a real-valued effective time. 
In the second subsection the relevant matrix elements of the effective time evolution operator entering these integral representations are evaluated algebraically with the help of a particular $so(2,1)$ subalgebra of the $so(4,2)$ algebra. Aspects relevant for the numerical evaluation of hydrogenic Lamb shifts and radiative decay rates on the basis of these integral representation are discussed in the subsequent sections. Sec. \ref{Eval1} discusses how the effective time integrations of the integral representations introduced in Sec. \ref{LieR} can partly be replaced by imaginary-valued effective time integrations with the help of Cauchy's integral theorem in order to avoid numerical problems with possibly rapidly oscillating integrands.  In Sec. \ref{Evaldipole} it is briefly summarized how the dipole approximation modifies the results of the previous sections. In order to exemplify the usefulness of the presented algebraic approach numerical results for hydrogenic Lamb shifts and radiative decay rates obtained from the integral representations introduced in Sec.\ref{LieR} are presented in Sec. \ref{Evalnum}. In order to facilitate the reading mathematical details, which are important for a self contained presentation but may distract somewhat from the main physical focus of this paper, are shifted to three appendices. Thus, appendix \ref{so42} briefly summarizes basic representation independent properties of the Lie algebra $so(4,2)$. Appendix \ref{sumrule1} derives the sum rule previously discussed by Maclay \cite{Maclay}, which is required for the integral representations of Sec.\ref{LieR}, and generalizes it to arbitrary representations. Finally, in appendix \ref{efftime1} the group theoretical derivation of the matrix elements of the effective time evolution operator entering the integral representations of Sec. \ref{LieR} is presented.

\section{Non-relativistic hydrogenic Lamb shifts and radiative decay rates\label{NRL}}
In this section basic facts are summarized about Lamb shifts and the radiative spontaneous decay rates of energy eigenstates of hydrogen-like ions obtainable in the framework of non-relativistic quantum electrodynamics in lowest order perturbation theory in the fine-structure constant. The non-relativistic formulation of quantum electrodynamics used here forms the basis of many quantum optical investigations \cite{CohenTannoudji}. The subsequent discussion focusses on non-relativistic electron velocities but, contrary to treatments based on the dipole approximation, it  takes into account retardation effects originating from the interaction of the hydrogenic electron with the electromagnetic field.

Neglecting the spin degrees of freedom, in the Schr\"odinger picture a simple non-relativistic model describing the interaction of the electron of a hydrogen-like ion with nuclear charge number $Z$ with the quantized radiation field in the Coulomb gauge is defined by the Hamiltonian
\begin{eqnarray}
H &=& :\frac{\left({\bf P}- e{\bf A}({\bf X})\right)^2}{2m}: -\frac{Ze^2}{4\pi \epsilon_0 |{\bf X}|} + H_F.
\label{H1}
\end{eqnarray}
 Thereby,
\begin{eqnarray}
H_A&=& \frac{{\bf P}^2}{2m} -\frac{Ze^2}{4\pi \epsilon_0 |{\bf X}|}
\label{HA}
\end{eqnarray}
is the Hamiltonian of the free hydrogen-like ion in the infinite nuclear mass limit with the bare electron mass $m$ and electron charge $e$, and
\begin{eqnarray}
H_F &=& \sum_{{\bf k},\sigma} \hbar \omega_{{\bf k}}a^{\dagger}_{{\bf k},\sigma}a_{{\bf k},\sigma} 
\end{eqnarray}
is the Hamitonian of the free transversal electromagnetic field with the dispersion relation $\omega_{{\bf k}}=c |{\bf k}|$ of the photonic field modes characterized by $({\bf k},\sigma)$.($c$ is the speed of light in vacuum,  $\omega_{{\bf k}}$ and ${\bf k}$ are frequencies and wave vectors of the modes, and the index $\sigma\in \{1,2\}$ characterizes the photon's helicity.)
The Cartesian components of the canonically conjugated momentum and position operators ${\bf P}_m$ and ${\bf X}_m$ of the electron fulfil the Heisenberg commutation relations $[{\bf P}_m,{\bf X}_n]= -i\hbar \delta_{m,n}$ with the reduced Planck constant $\hbar$. The permittivity of free space is denoted by $\epsilon_0$. 
In (\ref{H1}) it is assumed that relativistic effects originating from the dynamics of the electron are negligible, because its possible velocities are small in comparison with the velocity of light in vacuum.

In the Coulomb gauge the coupling between the transverse electromagnetic field and the electron is described by the transverse vector potential whose mode decomposition is given by 
\begin{eqnarray}
{\bf A}({\bf x}) &=&\sum_{{\bf k},\sigma}\sqrt{\frac{\hbar}{2\epsilon_0\omega_{{\bf k}}}}\left(
\frac{e^{i{\bf k}\cdot{\bf X}}}{\sqrt{V}} e_{{\bf k},\sigma} a_{{\bf k},\sigma} +
\frac{e^{-i{\bf k}\cdot{\bf X}}}{\sqrt{V}} e^{*}_{{\bf k},\sigma} a^{\dagger}_{{\bf k},\sigma}
\right). 
\label{A}
\end{eqnarray}
Thereby, the orthonormal unit vectors $e_{{\bf k},\sigma}$ with $\sigma \in \{1,2\}$ describe the photon's two helicity degrees of freedom and have to be chosen so that they are orthogonal to the wave vectors ${{\bf k}} = 2\pi{{\bf n}}/L$ with $0\neq {{\bf n}}\in {\mathbb Z}^3$. The quantization volume is denoted by $V=L^3$ and eventually tends to infinity in the continuum limit $L\to\infty$ so that
$(1/V)\sum_{{\bf k}}\longrightarrow (2\pi)^{-3}\int_{{\mathbb R}^3}d^3 k$. The bosonic annihilation and creation operators of the transverse modes of the electromagnetic field
 $a_{{\bf k},\sigma}$ and $a^{\dagger}_{{\bf k},\sigma}$ fulfil the characteristic commutation relations
 $[a_{{\bf k},\sigma}, a^{\dagger}_{{\bf k}',\sigma'}]=\delta_{{\bf k},{\bf k}'}\delta_{\sigma,\sigma'}$.
 The symbol $::$ in Eq.(\ref{H1}) denotes the normal ordering of the bosonic field operators.
 
 Non-relativistic models similar the one defined by Eq.(\ref{H1}) are typical starting points of quantum optical investigations describing the interaction of the quantized electromagnetic field with atoms \cite{CohenTannoudji}.  Typically, in these models the material non-spinorial degrees of freedom are described by canonically conjugated momentum and position operators and the independent degrees of freedom of the transverse electromagnetic field are described by bosonic field operators. 
The ground state of the transverse electromagnetic field is the vacuum state $|0\rangle$ with the characteristic property $a_{{\bf k},\sigma}|0\rangle = 0$ for all modes $({\bf k}, \sigma)$. The discrete energy eigenstates $|N L M\rangle$ of the bare hydrogenic Hamiltonian $H_A$ with well defined angular momenta $L\in {\mathbb N}_0$, $-L\leq M\in {\mathbb Z} \leq L$ and energy $\tilde{E}_N$ with $L+1\leq N\in {\mathbb N}$ are given by the Rydberg formula $\tilde{E}_N=-mc^2(Z\alpha_0)^2/(2N^2)$ with the dimensionless fine-structure constant $\alpha_0 = e^2/(4\pi \epsilon_0 \hbar c)$. 

The coupling between the hydrogen-like ion and the transverse electromagnetic field, whose strength is characterized by the fine-structure constant $\alpha_0$, modifies the ground state of the interacting Hamiltonian of (\ref{H1}). In addition, the excited states of the free (bare) Hamiltonian $H_A$ give rise to resonances, which can be characterized perturbatively by complex-valued energies $\tilde{E}_{N L M}$ whose real parts describe their energy positions and whose imaginary parts characterize their widths. In lowest (second) order perturbation theory in the transverse vector potential these complex-valued energies are given by
\begin{eqnarray}
{E}_{N L M} &=& \tilde{E}_N +\Delta \tilde{E}_{N L M},\nonumber\\
\Delta \tilde{E}_{N L M} &=&\frac{e^2\hbar}{2 m^2\epsilon_0 V}\lim_{\epsilon \to 0}
\sum_{{\bf k},\sigma}\frac{1}{\omega_{{\bf k}}}
\langle N L M |{\bf P}\cdot e_{{\bf k}, \sigma}
\left(
e^{i{\bf k}\cdot{\bf X}}\left(
\tilde{E}_N - H_A - \hbar \omega_{{\bf k}}+i\epsilon
\right)
e^{-i{\bf k}\cdot{\bf X}}
\right)^{-1}{\bf P}\cdot 
e^*_{{\bf k}, \sigma}
|N L M \rangle.
\end{eqnarray}
With the help of the shift property $e^{i{\bf k}\cdot{\bf X}}{\bf P}e^{-i{\bf k}\cdot{\bf X}} = {\bf P} -\hbar {\bf k}$, effects of the photon recoil on the hydrogenic electron can be taken into account. Neglecting these recoil effects would correspond to the dipole approximation. In particular, in the continuum limit and for non-relativistic electronic velocities
 we obtain the expression
\begin{eqnarray}
\Delta \tilde{E}_{N L M} &=&\frac{4\alpha_0}{3\pi m}\lim_{\epsilon \to 0}
\int_0^{x_>} dx 
\langle N L M|
{\bf P}\cdot \frac{\tilde{E}_N - H_A}{2mc^2 (1+x)}\left(
\frac{\tilde{E}_N - H_A}{2mc^2}-x(1+x) + i \epsilon
\right)^{-1}{\bf P}
| N L M \rangle -\nonumber\\
&&
\frac{4\alpha_0}{3\pi m}
\langle N L M | {\bf P}^2  | N L M \rangle \int_0^{x_>} \frac{dx}{1+x}
\label{DEdiv}
\end{eqnarray}
with the dimensionless photon energies $x= \hbar \omega_{{\bf k}}/(2mc^2)$ and the corresponding ultraviolet photonic cut-off frequency $x_>$. Formally, in this approximation terms of the order of $\mid {\bf P}\mid/(mc)$ are considered to be negligibly small in comparison with unity.
Contrary to the first term of (\ref{DEdiv}), its second term is logarithmically divergent in the ultraviolet limit of infinite photon frequencies, i.e. $x_> \to \infty$. This latter term can be interpreted as a lowest order contribution to a renormalization of the electron mass. Within such a mass renormalization framework, to lowest order in the fine-structure constant the physically observable electronic mass $m_e$ is related to its bare (photon frequency cut-off dependent) mass $m$ by the relation
\begin{eqnarray}
m_e &=& \frac{m}{1-\frac{8\alpha_0}{3\pi}\ln(1+x_>)}.
\end{eqnarray}
It is apparent that for a given finite physically observable mass $m_e$ such a mass renormalization is necessarily only possible in a consistent way as long as the frequency cut-off is chosen so that the bare mass does not vanish, i.e. $x_> < e^{3\pi/(8\alpha_0)}-1$.
As a result of the coupling to the transverse electromagnetic field modes in lowest order perturbation theory in the fine-structure constant $\alpha_0$ the hydrogen-like ion can be described by an effective Hamiltonian whose complex-valued eigenenergies are given by
\begin{eqnarray}
E_{NLM} &=& E_N + \Delta E_{NLM},\label{shift}\\
\Delta E_{N L M} &=&\frac{4\alpha_0}{3\pi m_e}\lim_{\epsilon \to 0}
\int_0^{\infty} dx 
\langle N L M|
{\bf P}\cdot \frac{E_N - H_A}{2m_ec^2 (1+x)}\left(
\frac{E_N - H_A}{2m_ec^2}-x(1+x) + i \epsilon
\right)^{-1}{\bf P}
| N L M \rangle = \nonumber\\
&&\Delta E_{NLM}^{{\rm Lamb}} - i \hbar \frac{\Gamma_{NLM}}{2}
\nonumber
\end{eqnarray}
with the Rydberg energies $E_N = -E_0/(2N^2)$  being determined by the physically observable electron mass $m_e$ and its corresponding atomic unit of energy $E_0 = m_e c^2(Z\alpha_0)^2$. 
Within this non-relativistic framework the real-valued part of $\Delta E_{NLM}$, i.e. $\Delta E_{NML}^{{\rm Lamb}}$, is the Lamb shift of the hydrogenic state $|NLM\rangle$ and $\Gamma_{NLM}$ is the radiative decay rate originating from the spontaneous emission of photons irrespective of their helicity.
It is apparent that within this approximation the ground state of this effective Hamiltonian is stable, i.e. $\Gamma_{100} = 0$.

It is of interest to compare non-relativistic calculations of Lamb shifts and radiative decay rates of hydrogen-like states based on (\ref{shift}) with the original treatment by Bethe \cite{Bethe}. This latter treatment is additionally based on the dipole approximation in which in the transverse vector potential (\ref{A}) the dependence of the photonic transverse modes on the position of the electron is neglected. In our treatment this approximation would correspond to the formal replacement $x(1+x) \to x$ in (\ref{DEdiv}). This changes the influence of the vacuum fluctuation of the transverse electromagnetic field modes with energies larger than the electron's rest mass energy $2m_e c^2$ on the electronic dynamics significantly even in the non-relativistic limit of small electronic velocities. Apart from a different mass renormalization of the form $m_e = m/(1-8\alpha_0 x_>/(3\pi))$ this dipole approximation also leads to well known logarithmic divergences of the Lamb shifts in the limit of large photon frequencies, i.e. $x_> \to \infty$. However, as apparent from the integral representation of (\ref{shift}), without such a dipole approximation the hydrogenic Lamb shifts yield finite values even in the limit of arbitrarily high virtual photon frequencies.
This has been realized by Seke \cite {Seke,Seke1}. However, within the non-relativistic approach yielding (\ref{shift}) the relativistic quantum electrodynamic contributions originating from high photon frequencies are not taken into account perturbatively up to second order in the fine-structure constant in a systematic way \cite{Bethe}.

For the evaluation of hydrogenic Lamb shifts and radiative decay rates according to (\ref{shift}), which take into account effects of the Coulomb potential on the electron dynamics non-perturbatively, 
it is necessary to evaluate matrix elements of the hydrogenic resolvent operator for hydrogenic energy eigenstates and arbitrary virtual transverse photon energies. On the one hand evaluations based on a spectral representation of this resolvent in terms of hydrogenic energy eigenstates, for example, are typically complicated by the presence of the continuous part of the hydrogenic energy spectrum. 
On the other hand evaluations based on explicit forms of the hydrogenic Green's function, such as path integral representations of the radial Green's function \cite{Grosche}, are complicated by the additionally required evaluation of the relevant matrix elements. Thus, it is worth investigating direct approaches for evaluating the Lamb shifts and radiative decay rates of (\ref{shift}) which exploit explicitly the already known high symmetry properties of the hydrogenic Hamiltonian, such as encoded in the $so(4,2)$ Lie algebra.

\section{The Lie algebra $so(4,2)$ and the non-relativistic hydrogenic Hamiltonian\label{LieH}}
In this section basic properties of a particular representation of the Lie algebra $so(4,2)$ are summarized which 
has been discussed recently in detail by Maclay \cite{Maclay} and forms the basis of our subsequent treatment. It
exhibits characteristic algebraic symmetry properties of the hydrogenic Hamiltonian. 
This representation offers the advantage that its basis only consists of hermitian operators with respect to the ordinary scalar product in the Hilbert space of the hydrogenic electron.
Alternative representations of this algebra, which do not share this particular property, have been used previously in various contexts including the
characterization of discrete energy eigenstates of hydrogen \cite{Englefield,Wybourne} or two-electron atoms \cite{Rau1990} or of the structure of two-qubit states \cite{Rau}.

The non-relativistic dynamics of a hydrogen-like ion  and its coupling to the electromagnetic field as described by the Hamiltonian (\ref{H1}) can be treated conveniently with the help of self-adjoint linear operators which constitute a representation of the Lie algebra $so(4,2)$. 
Starting from the quantum mechanical position and canonical momentum operators ${\bf X}$ and ${\bf P}$ the following dimensionless operators can be defined
\begin{eqnarray}
{\bf L} &=& {\bf X}\wedge {\bf P}/\hbar,\nonumber\\
{\bf A}(a) &=&\frac{1}{2a}\left( \frac{{\bf X}{\bf P}^2+{\bf P}^2{\bf X}}{2\hbar^2} - ({\bf X}\cdot{\bf P}){\bf P}/\hbar^2-{\bf P}({\bf P}\cdot {\bf X})/\hbar^2 - \frac{{\bf X}}{4r^2}  \right) -\frac{a{\bf X}}{2} ,\nonumber\\
{\bf G}(a) &=&\frac{1}{2a}\left( \frac{{\bf X}{\bf P}^2+{\bf P}^2{\bf X}}{2\hbar^2} - ({\bf X}\cdot{\bf P}){\bf P}/\hbar^2-{\bf P}({\bf P}\cdot {\bf X})/\hbar^2 - \frac{{\bf X}}{4r^2}  \right) +\frac{a{\bf X}}{2} ,\nonumber\\
G_4&=&\frac{1}{2}\left( {\bf X}\cdot {\bf P} + {\bf P}\cdot {\bf X}   \right)/\hbar,\nonumber\\
\Gamma_0(a) &=&\frac{1}{2}\left( \frac{\sqrt{r}{\bf P}^2 \sqrt{r}}{a\hbar^2}  + a r \right),\nonumber\\
{\bf \Gamma} &=& \sqrt{r} {\bf P} \sqrt{r}/\hbar,\nonumber\\
\Gamma_4(a) &=&\frac{1}{2}\left( \frac{\sqrt{r} {\bf P}^2 \sqrt{r}}{a\hbar^2}  - a r \right)
\label{so42hydrogen}
\end{eqnarray}
with $r^2 = {\bf X}^2$ and with $a > 0$ denoting an arbitrary constant with the physical dimension of a wave number. 
In view of the symmetric operator ordering all these operators are hermitian (symmetric)  with respect to the canonical scalar product $\langle \phi,\psi\rangle := \int_{{\mathbb R}^3}d^3x ~\phi^*({\bf x}) \psi({\bf x})$ in the Hilbert space $L_2(\mathbb{R}^3,\langle.,.\rangle$). On appropriately chosen dense subspaces of this Hilbert space these operators are also self-adjoint. Thereby, the operator ${\bf L} = \sum_{i=1,2,3}L_i {\bf e}_i$ with its three Cartesian components $L_i$ is the canonical (dimensionless) angular momentum operator (in units of $\hbar$) and generator of rotations. The three components of each of the dimensionless operators ${\bf A}(a), {\bf G}(a)$ or ${\bf \Gamma}$ transform as three-vectors and the operators $G_4, \Gamma_0(a)$ and $\Gamma_4(a)$ as scalars under rotations. The six Cartesian components of the operators ${\bf L}$ and ${\bf A}(a)$ form the generators of the subalgebra $so(4)\subset so(4,2)$ with the four components of the operator $({\bf G}(a),G_4)$ transforming  as the components of a four-vector with respect to the compact Lie group $SO(4)$ generated by $so(4)$. The operator ${\bf A}(a)$ is the quantum mechanical Laplace-Runge-Lenz-Pauli vector. The components of the operators ${\bf L}, {\bf A}(a), {\bf G}(a)$ and $G_4$ are the generators of the non-compact group $SO(4,1)$ with the combined operator $(\Gamma_0(a), {\bf \Gamma}, \Gamma_4(a))$ transforming as a five-vector under this group. Finally, the 15 components of the operators (\ref{so42hydrogen}) form an irreducible representation of the Lie algebra $so(4,2)$.
It fulfils the characteristic relations (cf. \cite{Maclay} and appendix \ref{so42})
\begin{eqnarray}
\Gamma^2(a) := \sum_{a=0}^4\Gamma_a(a) g_{aa}\Gamma_a(a) = -\Gamma_0^2(a) + {\bf \Gamma}^2+\Gamma_4^2(a) &=&1, \nonumber\\
G_4^2 + \Gamma_4^2(a) - \Gamma_0^2(a) + {\bf L}^2 &=&0,\nonumber\\
Q(a) := {\bf G}^2(a) - {\bf A}^2(a) + G_4^2 - {\bf L}^2 &=&2
\label{rel1}
\end{eqnarray}
and is characterized by the following values of its three independent Casimir operators
\begin{eqnarray}
I_1 &=& Q(a) + \Gamma^2(a) = 3,~~I_2 = 0,~~I_3 = 6(\Gamma^2(a))^2 - 24\Gamma^2(a) = -18.
\label{Casimir1}
\end{eqnarray}

Acting on the position and canonical momentum operators ${\bf X}$ and ${\bf P}$, the scalar operator $G_4$ generates scaling transformations, i.e.
\begin{eqnarray}
e^{i\lambda G_4}{\bf X}e^{-i\lambda G_4} &=& e^{\lambda}{\bf X},
\nonumber\\
e^{i\lambda G_4}{\bf P}e^{-i\lambda G_4} &=& e^{-\lambda}{\bf P}
\end{eqnarray}
with $\lambda \in {\mathbb R}$.
This is a consequence of the Heisenberg commutation relations. These scaling relations imply relations between equivalent representations of $so(4,2)$, which correspond to different values of $a$ according to the replacement $a\to a e^{\lambda}$. Thus, we find the relation
\begin{eqnarray}
e^{i\lambda G_4}\Gamma_0(a) e^{-i\lambda G_4} &=& \Gamma_0(ae^{\lambda}) = \Gamma_0 (a) \cosh \lambda -\Gamma_4(a) \sinh \lambda,
\label{scaling1}
\end{eqnarray}
for example.
Furthermore, the following important relation between the hydrogenic Hamiltonian $H_A$ and $\Gamma_0(a)$ can easily be found
\begin{eqnarray}
H_A - E &=& \frac{a\hbar^2}{m_e \sqrt{r}}\left(\Gamma_0(a) -\frac{a_1}{a} \right)\frac{1}{\sqrt{r}}.
\label{hydrogen+Gamma0}
\end{eqnarray}
It is valid for arbitrary values of $E = -(\hbar a)^2/(2m_e) < 0$
with $a_1 = Z\alpha_0 m_ec/\hbar=Z/a_0$, the Bohr radius $a_0$ and $a>0$. From this relation it becomes apparent that hydrogenic energy eigenstates of negative energies are closely related to eigenstates of $\Gamma_0(a)$ for arbitrary values of $a>0$. In particular, 
for the hydrogenic energy eigenstates $|NLM\rangle$ of energy $E_N = -m_ec^2(Z\alpha_0)^2/(2N^2)  < 0$ with $N\in {\mathbb N}$ and with definite angular momentum $L\in {\mathbb N}_0$, $N\geq L+1$, $-L\leq M\in {\mathbb Z} \leq L$ one finds the relation
\begin{eqnarray}
0 &=& (H_A - E_N) |NLM\rangle = \frac{a_1\hbar^2}{m_e N \sqrt{r}}\left(\Gamma_0(a_N) - N \right)\frac{1}{\sqrt{r}}|NLM\rangle = \frac{a_1^{3/2}\hbar^2}{m_e N^2\sqrt{r}}e^{i\lambda G_4}\left(
\Gamma_0(a) - N
\right)|N L M ; a)
\label{eigenvalue1}
\end{eqnarray}
with $e^{\lambda} = a_1/(a N)$ and 
with the normalized and rescaled states
\begin{eqnarray}
|NLM;a) &=& \sqrt{\frac{N^2}{a_1}}e^{-i\lambda G_4 }\frac{1}{\sqrt{r}}|NLM\rangle.
\label{eigenGamma0}
\end{eqnarray}
In the special case $a=a_1/N:=a_N$ this yields the relation $|NLM;a_N) = N/\sqrt{a_1 r}|NLM\rangle$.
That these rescaled states are normalized is a consequence of the virial theorem applied to the hydrogenic Hamiltonian $H_A$ implying the relation $\langle NLM| Ze^2/(4\pi\epsilon_0 r) |NLM\rangle = -2E_N$ \cite{CohenTannoudji1}.
Thus, for an arbitrary fixed value of $a>0$, according to (\ref{eigenGamma0}) each properly rescaled, normalized hydrogenic energy eigenstate $|NLM\rangle$ yields a normalized eigenstate $|NLM;a)$ of $\Gamma_0(a)$ with eigenvalue $N \in {\mathbb N}$. As $\Gamma_0(a)$ is self-adjoint and these eigenvalues exhaust its complete spectrum, for each fixed value of $a$ the states $|NLM;a)$ with $N\in {\mathbb N}$, $L\in {\mathbb N}_0$, $N\geq L+1$, $-L\leq M\in {\mathbb Z} \leq L$ constitute a purely discrete orthonormal basis in the Hilbert space of the hydrogenic electron.

\section{The Lie algebra $so(4,2)$ and non-relativistic hydrogenic Lamb shifts and radiative decay rates \label{LieR}}
Based on the particular representation of the Lie algebra $so(4,2)$ introduced in Sec. \ref{LieH} in this section integral representations are derived for the complex-valued energy shifts  (\ref{shift}) by which hydrogenic Lamb shifts and radiative decay rates of all energy eigenstates of definite angular momenta can be evaluated in a unified way. 

\subsection{General integral representations of the complex-valued hydrogenic energy shifts of (\ref{shift})\label{general}}

As apparent from (\ref{shift}), for the determination of Lamb shifts and radiative decay rates of bound states of a hydrogen-like ion with Hamiltonian $H_A$ one has to evaluate appropriate matrix elements of the retarded hydrogenic resolvent operator
\begin{eqnarray}
G_+(E) :=\left(H_A - E -i\epsilon \right)^{-1}
\end{eqnarray}
with $E\in {\mathbb R}$ and $\epsilon \to 0$ in the distributional sense. By a Laplace transformation this retarded resolvent is related to its associated time evolution operator $e^{-iH_A t/\hbar}$ by
\begin{eqnarray}
G_+(E) = \frac{i}{\hbar}\lim_{\epsilon \to 0}\int_0^{\infty}dt~e^{i(E+i\epsilon)t/\hbar}~e^{-iH_A t/\hbar}
\label{Laplace}
\end{eqnarray}
with the physical (Newtonian) time $t$. This relation between the retarded resolvent and its corresponding Hamiltonian $H_A$ is valid for any self-adjoint operator $H_A$ whose spectrum is necessarily real-valued.
Thus, the required matrix elements of the hydrogenic retarded resolvent operator can be represented by time integrations over corresponding matrix elements of the hydrogenic time evolution operator. Taking advantage of the symmetry properties of the hydrogenic Hamiltonian $H_A$ (cf. (\ref{so42hydrogen}),(\ref{hydrogen+Gamma0}), (\ref{eigenGamma0}))
one finds from (\ref{Laplace}) for $E = E_N - 2m_e c^2 x(1+x) = -(\hbar a)^2/(2m_e)$ the relation
\begin{eqnarray}
&&\langle NLM | {\bf P}\cdot \left( H_A - E - i\epsilon \right)^{-1}\cdot {\bf P}|NLM\rangle =
\frac{m_e \nu}{N^2}(NLM;a_N |
{\bf \Gamma} \cdot \left(\Gamma_0(a) - \nu -i\epsilon \right)^{-1}\cdot {\bf \Gamma} | NLM;a_N) =\nonumber\\
 &&i\frac{m_e\nu}{N^2}\int_0^{\infty}dT~e^{i(\nu+i\epsilon)T}(NLM;a_N|{\bf \Gamma}\cdot e^{-i\Gamma_0(a)T}{\bf \Gamma}|NLM;a_N)
 \label{complexshiftso42}
\end{eqnarray}
with $a_1 = Z\alpha_0 m_ec/\hbar, a_1/a=\nu$, 
with the effective dimensionless time-like variable $T$ and the associated effective time evolution operator $e^{-i\Gamma_0(a) T}$ being an element of the Lie group $SO(4,2)$. 
As shown in \cite{Maclay} and in appendix \ref{sumrule1}
the particular representation (\ref{so42hydrogen}) of the Lie algebra $so(4,2)$ implies the general sum rule 
\begin{eqnarray}
&&i\sum_{i=0}^4 g_{ii}\Gamma_i(a) \int_0^{\infty} dT ~e^{-in^T\Gamma (a) T}e^{i(\nu+i\epsilon) T}
\Gamma_i(a) 
=2i\int_0^{\infty}dT ~e^{i(\nu+i\epsilon) T}
 \frac{d^2}{dT^2}\left(\sin^2(T/2)
e^{-in^T\Gamma (a) T}\right) = \nonumber\\
&&
\sum_{i=0}^4 g_{ii}\Gamma_i(a) (n^T\Gamma (a) - \nu - i\epsilon)^{-1} \Gamma_i(a)
\label{sumrule}
\end{eqnarray}
with
$1=g_{11}=g_{22}=g_{33}=g_{44}=-g_{00}$, $n^T \Gamma(a) = \sum_{j=0}^4 n_j \Gamma_j(a), ~n\in {\mathbb R}^5$ and $\mid -n_0^2 + \sum_{j=1}^4 n_j^2\mid = 1$.
This sum rule not only relies on the commutation relations of the Lie algebra $so(4,2)$ but also on the particular representation of (\ref{so42hydrogen}) and in particular on the relation $\Gamma^2(a)= 1$ of (\ref{rel1}). Inserting (\ref{sumrule}) into (\ref{complexshiftso42}) with $\nu = Ne^{-\Phi}=a_1/a$, $a_N= a_1/N$ and $n = (\cosh\Phi,0,0,0,-\sinh\Phi)^T$ we obtain in the limit $\epsilon \to 0$
\begin{eqnarray}
&&\langle NLM | {\bf P}\cdot \left( H_A - E - i\epsilon \right)^{-1} {\bf P}|NLM\rangle = \Delta + \nonumber\\
&&
2i\frac{m_e e^{-\Phi}}{N}
\int_0^{\infty}dT ~e^{i(N e^{-\Phi}+i\epsilon) T}
 \frac{d^2}{dT^2}\left(\sin^2(T/2)
(NLM;a_N|e^{-i\Gamma (a_Ne^{\Phi}) T}|NM;a_N)\right)
\label{help1}
\end{eqnarray}
with (cf. appendix \ref{sumrule1})
\begin{eqnarray}
\Delta &=&
\frac{m_e\nu}{N^2}
(NLM;a_N|\left(\Gamma_0(a_N) \left(\Gamma(a_N e^{\Phi}) - Ne^{-\Phi} -i\epsilon\right)^{-1}\Gamma_0(a_N) ~- \right.\nonumber\\
&&\left.\Gamma_4(a_N) \left(\Gamma(a_N e^{\Phi}) - Ne^{-\Phi} -i\epsilon \right)^{-1}\Gamma_4(a_N)\right)|NLM;a_N)=
\frac{\langle NLM| {\bf P}^2|NLM\rangle}{E_N-E}
\label{Delta1}
\end{eqnarray}
contributing to the mass renormalization.
Inserting (\ref{complexshiftso42}) and (\ref{help1})  into (\ref{DEdiv}) the (renormalized) complex-valued energy shift $\Delta E_{NLM}$ of (\ref{shift}) can be rewritten in the form
\begin{eqnarray}
\Delta E_{NLM} &=&-i\frac{4m_e c^2 \alpha_0 (Z\alpha_0)^2}{3\pi N^3}\int_0^{\infty}d\Phi e^{\Phi}
\frac{-1+\sqrt{1+2(Z\alpha_0/N)^2e^{\Phi}\sinh\Phi}}{
\sqrt{1+2(Z\alpha_0/N)^2e^{\Phi}\sinh\Phi}}
\lim_{\epsilon \to 0}
\int_0^{\infty}dT~e^{i(Ne^{-\Phi}+i\epsilon)T}
\frac{d^2 Q_{NLM}}{dT^2}(T,\Phi)\nonumber\\
\label{complexshiftfinal}
\end{eqnarray}
with 
\begin{eqnarray}
Q_{NLM}(T,\Phi) &=& 
\sin^2(T/2) 
(NLM;a_N| e^{-i\Gamma_0(a_N e^{\Phi}) T}|NLM;a_N)
\label{Q}
\end{eqnarray}
and $a_N=a_1/N, E_N = -(\hbar a_N)^2/(2m_e)$. In the subsequent subsection it will be demonstrated that for fixed values of $L$ and $M$ the matrix elements of the effective time evolution operator $(NLM;a_N| e^{-i\Gamma_0(a_N e^{\Phi}) T}|NLM;a_N)$ are closely related to infinite dimensional irreducible unitary representations of $SO^+(2,1)=SU(1,1)/{\mathbb Z}_2$ and can thus be evaluated by group theoretical methods.
Alternative expressions for the complex-valued energy shift can be obtained from (\ref{complexshiftfinal}) by partial integration with respect to the variable $T$, i.e.
\begin{eqnarray}
\Delta E_{NLM} &=&-\frac{4m_e c^2 \alpha_0 (Z\alpha_0)^2}{3\pi N^2}\int_0^{\infty}d\Phi
\frac{-1+\sqrt{1+2(Z\alpha_0/N)^2e^{\Phi}\sinh\Phi}}{
\sqrt{1+2(Z\alpha_0/N)^2e^{\Phi}\sinh\Phi}}\lim_{\epsilon \to 0}
\int_0^{\infty}dT~e^{i(Ne^{-\Phi}+i\epsilon)T}
\frac{dQ_{NLM}}{dT}(T,\Phi) =\nonumber\\
 \label{shiftnumerical}\\
&&i\frac{4m_e c^2 \alpha_0 (Z\alpha_0)^2}{3\pi N}\int_0^{\infty}d\Phi e^{-\Phi}
\frac{-1+\sqrt{1+2(Z\alpha_0/N)^2e^{\Phi}\sinh\Phi}}{
\sqrt{1+2(Z\alpha_0/N)^2e^{\Phi}\sinh\Phi}}\lim_{\epsilon \to 0}
\int_0^{\infty}dT~e^{i(Ne^{-\Phi}+i\epsilon)T}Q_{NLM}(T,\Phi).
\label{alternative}\nonumber
\end{eqnarray}

\subsection{Group theoretical evaluation of the effective time evolution matrix elements 
$(NLM;a_N| e^{-i\Gamma_0(a_N e^{\Phi}) T}|NLM;a_N)$}

From their definition (cf. (\ref{so42hydrogen}) it is apparent that the three hermitian operators $\left\{\Gamma_4(a_N),G_4,\Gamma_0(a_N)\right\}$ are a basis of the subalgebra $so(2,1)\simeq su(1,1) \subset so(4,2)$. Thus, renaming these operators by $J_1:=\Gamma_4(a_N), J_2:= G_4, J_3:= \Gamma_0(a_N)$ they fullfil the characteristic commutation relations
\begin{eqnarray}
\left[J_1,J_2  \right]&=&-iJ_3,~\left[J_1,J_3  \right] = -iJ_2,~\left[J_2,J_3  \right] = iJ_1
\label{comm}
\end{eqnarray}
with the Casimir operator $J^2 := J_1^2 + J_2^2 - J_3^2$. As apparent from (\ref{rel1}), for the particular representation of $so(4,2)$ defined by (\ref{so42hydrogen}) this Casimir operator is related to the squared angular momentum operator ${\bf L}^2$ by
\begin{eqnarray}
J^2 &=& -\bf{L}^2
\end{eqnarray}
so that the generators $J_1, J_2, J_3$ form a subalgebra of this $so(4,2)$ representation.
Thus, inequivalent representation spaces of this subalgebra $so(2,1)\simeq su(1,1)$ within the irreducible representation space of $so(4,2)$, i.e. within the physical Hilbert space of the hydrogenic electron, are characterized by different eigenvalues of the squared angular momentum operator $\bf{L}^2$. 

Starting from the commutation relations (\ref{comm})
all possible inequivalent irreducible continuous unitary representation spaces with orthonormal basis $\{|X m)\}$ of the associated Lie group $SU(1,1)$ can be constructed which diagonalize the commuting operators $J_3$ and $J^2$ simultaneously, i.e.
\begin{eqnarray}
J^2|X m) &=& X|X m),\nonumber \\
J_3|X m) &=& m|X m),
\label{simultaneous}
\end{eqnarray}
and for which the generator $J_3$ is bounded from below \cite{Englefield,Wybourne,Bargmann,Biedenharn,Chiribella2006} (cf. appendix \ref{efftime1}). The resulting possible eigenvalues are
$X=m_0(1-m_0)$ and $m=m_0+k$ with $k,2m_0\in {\mathbb N}_0$. Thus, consistency with the possible eigenvalues of $\Gamma_0(a_N)$ and the fact that the generators $J_1,J_2,J_3$ form a subalgebra of the $so(4,2)$ representation (\ref{so42hydrogen})
require the identifications $m_0=L+1 \in {\mathbb N}$ and $L+1\leq n = m_0+k\in {\mathbb N}$ for each fixed value of $L$ and $M\in {\mathbb Z}$ with $-L\leq M \leq L$ so that the rescaled orthonormal states
$\{|n L M;a_N)|~ n\in {\mathbb N}_0, n\geq L+1\}$ of (\ref{eigenGamma0}) are a basis of an irreducible unitary representation space of the group $SO^+(2,1)=SU(1,1)/{\mathbb Z}_2$ \cite{Bargmann,Biedenharn}. Thus, $SU(1,1)$ is the double cover of the identity-connected component of $SO(2,1)$. This connection with the Lie group $SU(1,1)$  implies that basic representation independent group properties of $SU(1,1)$ enable an evaluation of 
the matrix elements of the effective time evolution operator entering (\ref{Q}) and (\ref{shiftnumerical}). 
In particular,  it is demonstrated in detail in appendix \ref{efftime1}
that taking advantage of a well known Baker-Campbell-Hausdorff relation for the group $SU(1,1)$ \cite{Chiribella2006,Puri} yields the final result
\begin{eqnarray}
(NLM;a_N| e^{-i\Gamma_0(a_N e^{\Phi}) T}|NLM;a_N) &=& (f(T,\phi))^{-2N}~_2F_1(L+1-N,-L-N;1;1-\mid f(T,\Phi)\mid ^2) =\nonumber\\
&&e^{-i2N\chi(T,\Phi)}(1-z(T,\Phi))^LP_{N+L}^{(0,-1-2L)}\left(\frac{1+z(T,\Phi)}{1-z(T,\Phi)}\right),\nonumber\\
f(T,\Phi)&=& |f(T,\Phi)|e^{i\chi(T,\Phi)}=\cos(T/2)+i\sin(T/2) \cosh\Phi,\nonumber\\
z(T,\Phi) &=& 1-|f(T,\Phi)|^2 = -\sin^2(T/2)\sinh^2\Phi
\label{hyper}
\end{eqnarray}
with the hypergeometric functions $_2F_1(a,b;c;z)$ and the Jacobi polynomials $P_n^{(\alpha,\beta)}(w)$. In order to establish the relation between these hypergeometric functions and the Jacobi polynomials  we have used the identity
$_2F_1(L+1-N,-L-N;1,z) = (1-z)^{L+N}~_2F_1(-N-L,N-L;1;z/(z-1) = P_{N+L}^{(0,-1-2L)}((1+z)/(1-z))$ \cite{AS}.

\section{Evaluation of hydrogenic Lamb shifts and radiative decay rates \label{Eval1}}

For numerical evaluations of hydrogenic Lamb shifts and radiative decay rates based on 
the results of the previous section it is convenient to circumvent potential problems arising from the oscillatory $T$-dependence of the integrands of (\ref{complexshiftfinal}) and (\ref{shiftnumerical}). For large values of the principal quantum number $N$ such oscillations may cause significant numerical problems. For holomorphic functions problems arising from rapidly oscillating integrands can be circumvented with the help of Cauchy's integral theorem by deforming the relevant integration path in appropriate ways. In particular, in the case of the $T$-integrations involved in (\ref{complexshiftfinal}) and (\ref{shiftnumerical}) this can be achieved by noting that $Q_{NLM}(T,\Phi)$ of (\ref{Q}) is a holomorphic function of $T$ in the lower complex $T$-plane and that it involves both exponentially increasing and exponentially decreasing functions of $T$ in this region. Thus, it is convenient to separate these two types of contributions
by expanding $Q_{NLM}(T,\Phi)$ into a power series in the exponential functions $e^{-inT}$, i.e.
\begin{eqnarray}
Q_{NLM}(T,\Phi) &=& \Tilde{Q}_{NLM}(T,\Phi) + \sum_{n=L}^{N-1} e^{-inT} R_n^{(N,L)}(\cosh \Phi)
\label{Qseries}
\end{eqnarray}
with
\begin{eqnarray}
R_n^{(N,L)}(\cosh \Phi) &=&
\frac{1}{2}\mid (NLM;a_N|e^{i\Phi G_4}|nLM;a_N)\mid^2 -
\frac{1}{4}\mid (NLM;a_N|e^{i\Phi G_4}|(n+1) LM;a_N)\mid^2 -\nonumber\\
&&\frac{1}{4}\mid (NLM;a_N|e^{i\Phi G_4}|(n-1) LM;a_N)\mid^2,
\label{Rn}
\end{eqnarray}
so that $\tilde{Q}(T,\phi)$ contains all exponential terms with $n\geq N$. The matrix elements involved in (\ref{Qseries}) can be obtained directly from (\ref{finitetrafo}) and (\ref{scaling2}) of appendix \ref{efftime1} or alternatively by a power series expansion with respect to the variable $e^{-iT}$ with the help of algebraic programs, such as Mathematica \cite{Mathematica}. Thus, using contour integration and noting that the contribution of $\tilde{Q}_{NLM}(T,\Phi)$ vanishes along a semicircle at infinity in the lower complex $T$-plane, (\ref{shiftnumerical}) can be rewritten in the equivalent form
\begin{eqnarray}
\Delta E_{NLM} &=&-\frac{4m_e c^2 \alpha_0 (Z\alpha_0)^2}{3\pi N^2}\int_0^{\infty}d\Phi
\frac{-1+\sqrt{1+2(Z\alpha_0/N)^2e^{\Phi}\sinh\Phi}}{
\sqrt{1+2(Z\alpha_0/N)^2e^{\Phi}\sinh\Phi}}\times\nonumber\\
&&\left(
\int_0^{\infty}d\tau~e^{\tau Ne^{-\Phi}}
\frac{d\tilde{Q}_{NLM}}{d\tau}(T=-i\tau,\Phi) +\sum_{n=L}^{N-1}nR_n^{(N,L)}(\cosh\Phi)\left(
\frac{\cal{P}}{Ne^{-\Phi}-n}-i\pi\delta(Ne^{-\Phi}-n)
\right)
\right)
\label{shiftseparate}
\end{eqnarray}
with ${\cal P}$ indicating that in the principal value of the $\Phi$-integration has to be taken. 
Analogous expressions can easily be obtained for the other equivalent representations of $\Delta E_{NML}$ in (\ref{complexshiftfinal}) or (\ref{shiftnumerical}). 

The last term of (\ref{shiftseparate}) involving the $\delta$-distribution is purely imaginary and yields the total radiative decay rate $\Gamma_{NLM}$ of the hydrogenic energy eigenstate $|NLM\rangle$ to all radiatively coupled lower lying energy eigenstates with principal quantum numbers $1\leq n\leq N-1$. The partial radiative decay rate due to radiative transitions from the hydrogenic state $|NLM\rangle$ to energetically lower lying energy eigenstates with a particular principal quantum number $n$ (irrespective of the helicity of the spontaneously emitted photon) is given by
\begin{eqnarray}
\Gamma^{(N L M)}_n &=&-2\pi \frac{4\alpha_0}{3\pi N^2}
{R^{(N L)}_n} (\cosh \Phi_0)\frac{-1+\sqrt{1+2(Z\alpha_0/N)^2 e^{\Phi_0}\sinh\Phi_0}}{\sqrt{1+2(Z\alpha_0/N)^2 e^{\Phi_0}\sinh\Phi_0}}
\left(\frac{m_ec^2(Z\alpha_0)^2}{\hbar}\right)
\label{Gammaseparate}
\end{eqnarray}
with $e^{-\Phi_0} = n/N$. Correspondingly, for $N\geq 2$ the total spontaneous decay rate $\Gamma_{NLM}$ is given by
\begin{eqnarray}
\Gamma_{NLM} &=& \sum_{1\leq n=L}^{N-1}\Gamma^{(N L M)}_n
\end{eqnarray}
and the hydrogenic ground state $|100\rangle$ is stable. It should be mentioned that the corresponding radiative decay rates in the dipole approximation are obtained in the limit $(Z\alpha_0/N)^2 e^{\Phi_0}\sinh\Phi_0 \ll 1$.

The terms of (\ref{shiftseparate}) not involving the $\delta$-distribution are real-valued and yield the Lamb shift of the hydrogenic quantum state $|NLM\rangle$. Again, the dipole approximation corresponds to the limit $(Z\alpha_0/N)^2 e^{\Phi_>}\sinh\Phi_> \ll 1$ and implies that in this limit the hydrogenic Lamb shifts depend on the chosen frequency cut-off characterized by  $\Phi_>$. In this limit and in the special case of $N=1$ (\ref{shiftseparate}) reduces to the recently presented result of Maclay \cite{Maclay}.

\section{Complex-valued hydrogenic energy shifts and the dipole approximation \label{Evaldipole}}
The results in the dipole approximation are obtained from (\ref{complexshiftfinal}) and (\ref{shiftnumerical}) by neglecting the recoil effects affecting the energy change of the hydrogenic electron by virtual photon emission. Formally  this corresponds to the replacements
\begin{eqnarray}
E_{N} - E &=&2m_ec^2x(x+1) \longrightarrow~2m_ec^2x,\nonumber\\
x = \frac{\hbar \omega}{2m_ec^2} &=&\frac{1}{2}\left(-1+\sqrt{1+2(\frac{Z\alpha_0}{N})^2e^{\Phi}\sinh\Phi} \right) \longrightarrow~\frac{1}{4}(\frac{Z\alpha_0}{N})^2(e^{2\Phi}-1).
\label{dipole1}
\end{eqnarray}
As long as the dynamics of the hydrogenic electron is described non-relativistically, consistency with the dipole approximation requires that the energies of (virtual) photons should be restricted by an  upper cut-off energy $\hbar \omega_>$ large in comparison with the binding energy $E_0/2 = Z^2 {\rm Ryd} = m_ec^2(Z\alpha_0)^2/2$
 of the hydrogen-like ion, but small in comparison with the electronic rest mass energy $m_ec^2$. Thus, for $Z\alpha_0 \ll 1$ it should be of the order of $ m_ec^2 (Z\alpha_0)/2$. Contributions to the Lamb shift of (virtual) photon frequency larger than $\omega_>$ can be taken into account perturbatively by relativistic quantum electrodynamics. Both contributions taken together yield a cut-off independent Lamb shift. As a consequence, for a hydrogenic energy eigenstate  with principal quantum number $N$ and angular momentum $L$ in lowest order perturbation theory in the fine structure constant the Lamb shift is given by
\cite{BS,Huff,IZ}
\begin{eqnarray}
\Delta E_{NLJ} &=&
\frac{8\alpha_0^3 Z^4}{3\pi N^3}\frac{m_ec^2\alpha_0^2}{2}\times
\left\{
\begin{array}{ll}
\frac{19}{30}-\gamma(N,L) -2\ln(Z\alpha_0)&{\rm for}~L=0,~J=1/2\\
\frac{3c_{L,J}}{8(2L+1)}-\gamma(N,L)&{\rm for}~L\neq 0
\end{array}
\right.,\nonumber\\
c_{L,J} &=&\left\{
\begin{array}{cc}
(L+1)^{-1}&~J=L+1/2\\
-L^{-1}&~J=L-1/2.
\end{array}
\right.
\label{dipoleshift}
\end{eqnarray}
Thereby, the cut-off independent Bethe logarithm \cite{BS,Huff}
\begin{eqnarray}
\gamma(N,L) &=&  \lim_{\frac{\hbar\omega_>}{2m_ec^2}\to \infty}\left\{-\frac{3\pi N^3\Delta E^>_{NL}}{8\alpha_0^3Z^4(m_ec^2\alpha_0^2/2)}+\delta_{L,0}
\left(
\ln\frac{\hbar \omega_>}{m_ec^2/2}
-2 \ln(Z\alpha_0)\right)
\right\}
\label{Bethelogarithm}
\end{eqnarray}
characterizes the non-relativistic atomic contribution to the Lamb shift in terms of a characteristic mean excitation energy $\langle E_{NL}\rangle = (Z^2m_ec^2\alpha_0^2/2)\exp{\gamma(N,L)}$. The physical significance of the other  contributions entering (\ref{dipoleshift}) is of purely relativistic quantum electrodynamic origin and is discussed in detail in Ref. \cite{BS}, for example.
In the definition of the Bethe logarithm (\ref{Bethelogarithm}) the quantity $\Delta E^>_{NL}$ denotes the Lamb shift in the dipole approximation (cf. \ref{dipole1}) as obtained from (\ref{shiftseparate}) with a cut-off frequency, i.e. $x_>=\hbar \omega_>/(2m_ec^2) \gg 1$ \cite{Huff}.

\section{Numerical results\label{Evalnum}}
In order to exemplify the usefulness of the group theoretical integral representations of Sec. \ref{LieR} in this section numerical results for Lamb shifts and radiative decay rates of hydrogenic energy eigenstates with well defined angular momenta are presented which transcend the dipole approximation and can easily be obtained on the basis of (\ref{shiftseparate}) and (\ref{Gammaseparate}) with the help of standard algebraic and numerical routines, such as available in Mathematica \cite{Mathematica}. For the sake of comparison also corresponding results are presented in the dipole approximation. The numerical values of the required physical constants have been taken from Ref. \cite{Constants}.

\begin{table}[H]
\begin{center}
\begin{tabular}{|c|c|c|}
\hline
$(N,L,n)$& $\Delta E_{N L M}$&$\Gamma_n^{(N L M)}$\\
\hline
(1,0,1)&7936,29&0\\
\hline
(2,0,1)&1015,40&0\\
(2,1,1)&4,09715&626,813\\
\hline
(3,0,1)&302,626&0\\
(3,0,2)&&6,31698\\
(3,1,1)&1,53944&167,338\\
(3,1,2)&&22,4604\\
\hline
(4,0,1)&127,993&0\\
(4,0,2)&&2,57953\\
(4,0,3)&&1,83642\\
(4,1,1)&0,713471&68,2212\\
(4,1,2)&&9,67325\\
(4,1,3)&&3,41451\\
\hline
\end{tabular}
\caption{Lamb shifts $\Delta E_{N L M}$ of hydrogenic energy eigenstates with $Z=1$, principal quantum numbers $N$ and angular momentum quantum numbers $L\in\{0,1\}$ (in units  of $10^6$ Hz) and their corresponding radiative decay rates $\Gamma_n^{(N L M)}$ to all lower lying energy eigenstates with principal quantum numbers $n$ (in units of $10^6 s^{-1}$) according to (\ref{shiftseparate}) and (\ref{Gammaseparate}) without dipole approximation.}
\label{Nondipole}
\end{center}
\end{table}

Table \ref{Nondipole} depicts
Lamb shifts $\Delta E_{N L M}$ of hydrogenic $s$- and $p$-states with $Z=1$ and with principal quantum numbers $N\in \{1,2,3,4\}$ in units  of $10^6$ Hz. In addition, their corresponding radiative decay rates $\Gamma_n^{(N L M)}$ to all lower lying energy eigenstates with principal quantum numbers $n < N$  are shown 
in units of $10^6 s^{-1}$. In these radiative decay rates the helicity of the spontaneously emitted photon is not observed.  These quantities do not involve the dipole approximation and have been evaluated from (\ref{shiftseparate}) and (\ref{Gammaseparate})
in a straight forward way with the help of algebraic and numerical routines of Mathematica \cite{Mathematica}. These Lamb shifts without the dipole approximation are in good agreement with the previous results of Seke and M\"odritsch \cite{Seke,Seke1} for $N=1,2$, for example.

\begin{table}[H]
\begin{center}
\begin{tabular}{|c|c|c|c|c|c|}
\hline
$(N,n)$& $\Delta E_{N L=0 J=1/2}$&$\tilde{\Delta} E_{N L=0 J=1/2}$&$\Gamma_n^{(N L=0 M)}$&$\gamma(N,L=0)$&
$\langle E_{NL=0}\rangle$\\
\hline
(1,1)&8127,44&7440,18&0&2,98413&19,769\\
\hline
(2,1)&1039,31&953,402&0&2,81177&16,639\\
\hline
(3,1)&309,716&284,262&0&2,76767&15,921\\
(3,2)&&&6,31702&&\\
\hline
(4,1)&130,967&120,229&0&2,74965&15,637\\
(4,2)&     &     &2,57955&&\\
(4,3)&&&1,83642&&\\
\hline
\end{tabular}
\caption{Lamb shifts $\Delta E_{N L J}$ of hydrogenic energy eigenstates with $Z=1$, principal quantum numbers $N$ and angular momentum quantum number $L=0$ (in units  of $10^6$ Hz) and their corresponding radiative decay rates $\Gamma_n^{(N L=0 M)}$ to all lower lying energy eigenstates with principal quantum numbers $n$ (in units of $10^6 s^{-1}$) according to (\ref{dipoleshift}) based on the dipole approximation (cf (\ref{dipole1})): In addition, the Lamb shifts $\tilde{\Delta}E_{N L=0 J}$ without the relativistic quantum electrodynamic contribution proportional to $19/30$
(cf. (\ref{dipoleshift})),
 the Bethe-logarithms $\gamma(N, L=0)$ and their equivalent average energies $\langle E_{N L=0}\rangle$ as evaluated from (\ref{shiftseparate}) (in units of the Rydberg constant $E_0/2 = Z^2m_ec^2\alpha_0^2/2$) are shown.}
\label{Tabledipole1}
\end{center}
\end{table}

Whereas the perturbative Lamb shifts of Table \ref{Nondipole} depend only on the principal quantum number $N$ and the angular momentum quantum number $L$, according to (\ref{dipoleshift}) the corresponding results in the dipole approximation also depend on the total angular momentum $J=L\pm 1/2 \geq 0$ of the hydrogenic electron.  This is due to the fact that in 
(\ref{dipoleshift}) also relativistic quantum electrodynamic effects originating from high-energy virtual photons have been taken into account systematically in addition to the atomic contribution. This latter atomic contribution is characterized by the cut-off independent Bethe-parameter of (\ref{Bethelogarithm}). Table \ref{Tabledipole1} depicts
Lamb shifts $\Delta E_{N L=0 J}$ of hydrogenic $s$-states with principal quantum numbers $N\in \{1,2,3,4\}$ (in units  of $10^6$ Hz) and their corresponding radiative decay rates $\Gamma_n^{(N L=0 M)}$ to all lower lying energy eigenstates with principal quantum numbers $n < N$ (in units of $10^6 s^{-1}$).
These results are based on  (\ref{Gammaseparate}), (\ref{dipole1}) and (\ref{dipoleshift})
within the framework of the dipole approximation. In order to emphasize the significance of the relativistic contributions included in (\ref{dipoleshift}) also the 
Lamb shifts $\tilde{\Delta}E_{N L=0 J}$ without the relativistic quantum electrodynamic contributions proportional to $19/30$ are depicted. In addition, the corresponding values of
 the Bethe-logarithms $\gamma(N, L=0)$ and their equivalent average energies $\langle E_{N L=0}\rangle$ as evaluated from (\ref{shiftseparate}) (in units of the Rydberg constant $E_0/2 = Z^2m_ec^2\alpha_0^2/2$) are shown. The well-known weak dependences of the Bethe-logarithms and the corresponding average energies on the principal quantum number and the good agreement with previous results are apparent \cite{BS,IZ}.

\begin{table}[H]
\begin{center}
\begin{tabular}{|c|c|c|c|c|c|}
\hline
$(N,J,n)$& $\Delta E_{N L=1,J}$&$\tilde{\Delta} E_{N L=1}$&$\Gamma_n^{(N L=1)}$&$\gamma(N,L=1 M)$&
$\langle E_{N,L=1}\rangle$\\
\hline
(2,1/2,1)&-12,8840&4,07142&626,832&-0,0300156&0,97043\\
(2,3/2,1)&12,5492&&&&\\
\hline
(3,1/2,1)&-3,48894& 1,53491 &167,344&-0,0381905&0,96253\\
(3,3/2,1)&4,04683&&&&\\
(3,1/2,2)&-3,48894& 1,53491 &22,4605&-0,0381905&0,96253\\
(3,3/2,2)&4,04683&&&&\\
\hline
(4,1/2,1)&-1,40791&0,711522&68,2238&-0,0419642&0,95890\\
(4,3/2,1)&1,77124&&&&\\
(4,1/2,2)&-1,40791&0,711522&9,67333&-0,0419642&0,95890\\
(4,3/2,2)&1,77124&&&&\\
(4,1/2,3)&-1,40791&0,711522&3,41452&-0,0419642&0,95890\\
(4,3/2,3)&1,77124&&&&\\
\hline
\end{tabular}
\caption{Lamb shifts $\Delta E_{N L J}$ of hydrogenic energy eigenstates with $Z=1$, principal quantum numbers $N$ and angular momentum quantum number $L=1$ and $J=L\pm 1/2$ (in units  of $10^6$ Hz) and their corresponding radiative decay rates $\Gamma_n^{(N L=1 M)}$ to all lower lying energy eigenstates with principal quantum numbers $n$ (in units of $10^6 s^{-1}$) according to (\ref{dipoleshift}) based on the dipole approximation (cf (\ref{dipole1})): In addition, the Lamb shifts $\tilde{\Delta}E_{N L=1}$ without the relativistic quantum electrodynamic contributions proportional to $c_{L=1,J}$ 
(cf. (\ref{dipoleshift})),
 the Bethe-logarithms $\gamma(N, L=1)$ and the corresponding average energies $\langle E_{N, L=1}\rangle$ as evaluated from (\ref{shiftnumerical}) (in units of the Rydberg constant $E_0/2 = Z^2m_e^2c^2\alpha_0^2/2$) are shown.}
\label{Tabledipole2}
\end{center}
\end{table}

Table \ref{Tabledipole2} depicts
Lamb shifts $\Delta E_{N L=1 J}$ of hydrogenic $p$-states with principal quantum numbers $N\in \{1,2,3,4\}$ (in units  of $10^6$ Hz) and their corresponding radiative decay rates $\Gamma_n^{(N L=0 M)}$ to all lower lying energy eigenstates with principal quantum numbers $n < N$ (in units of $10^6 s^{-1}$).
Again these results are based on  (\ref{Gammaseparate}), (\ref{dipole1}) and (\ref{dipoleshift})
within the framework of the dipole approximation.
Also shown are  the 
Lamb shifts $\tilde{\Delta}E_{N L=1 J}$ without the relativistic quantum electrodynamic contributions proportional to $c_{L=1, J}$ according to (\ref{dipoleshift}) . In addition, the corresponding values of
 the Bethe-logarithms $\gamma(N, L=1)$ and their equivalent average energies $\langle E_{N L=1}\rangle$ as evaluated from (\ref{shiftseparate}) (in units of the Rydberg constant $E_0/2 = Z^2m_ec^2\alpha_0^2/2$) are shown. Besides the well-known weak dependence of the Bethe-logarithms and the corresponding average energies on the principal quantum number and the good agreement with previous results \cite{BS,IZ} it is also apparent that for $p$-states these quantities are significantly smaller than for $s$-states. 
 
 As far as the radiative decay rates in the dipole approximation are concerned, it is also apparent from Table \ref{Tabledipole2} that within the numerical accuracy of the presented results the decay rate of the hydrogenic $2p$-state ($N=2, L=1, Z=1$), for example, agrees with the exactly known value 
 \begin{eqnarray}
 \Gamma_1^{(2 1 M)} = -\frac{1}{4}R_1^{(2 1)}(5/4)
 \left(\frac{m_e c^2 \alpha_0(Z\alpha_0)^4}{\hbar}\right) = \left(\frac{2}{3}\right)^8
 \left(\frac{m_e c^2 \alpha_0 (Z\alpha_0)^4}{\hbar}\right)
 \label{Gammap}
 \end{eqnarray}
with $e^{-\Phi_0} = 1/2$ implying $\cosh\Phi_0 = 5/4$\cite{BS}. Analogously, in the dipole approximation the exact values of all other radiative decay rates can easily be determined in a straight forward way from (\ref{Gammaseparate}) with (\ref{dipole1}).
A particularly simple case arises for hydrogenic energy eigenstates of maximal angular momenta, i.e. $N=L+1$, for example, 
because $_2F_1(0,-2L-1;1;1-|f(T,\Phi)|^2) = 1$ in (\ref{hyper}). Thus, one obtains the radiative decay rates
\begin{eqnarray}
\Gamma_{N-1}^{(N~N-1 ~M)} &=& -\frac{4(2N-1)}{3N^4 (N-1)^2}R_{N-1}^{(N~ N-1)}(\cosh\Phi_0)
\left(\frac{m_e c^2 \alpha_0 (Z\alpha_0)^4}{\hbar}\right) = \nonumber\\
&&\frac{2}{3}\frac{N-1/2}{N^4(N-1)^2}\left(1+\frac{1}{4N(N-1)} \right)^{-2N}
\left(\frac{m_e c^2 \alpha_0 (Z\alpha_0)^4}{\hbar}\right)
\label{Gammacircular}
\end{eqnarray}
with $R_{N-1}^{(N~N-1)}(\cosh\Phi) = -1/(4\cosh^{4N}(\Phi/2))$ and $\cosh\Phi_0 = 1+1/(2N(N-1))$. For the special case of $N=2$ it reduces to (\ref{Gammap}).
For large principal quantum numbers, i.e.  $N\gg1$, it reduces to the corresponding semiclassical result of Marxer and Spruch \cite{MarxerSpruch} with corrections of the order of $N^{-2}$, i.e.
\begin{eqnarray}
\Gamma_{N-1}^{(N~N-1 ~M)} &=&\frac{2}{3}\frac{\left(1 + O(N^{-2})\right)}{N^4(N-1)}
\left(\frac{m_e c^2 \alpha_0 (Z\alpha_0)^4}{\hbar}\right).
\end{eqnarray}

\section{Summary and Conclusions}

An algebraic approach has been presented for evaluating Lamb shifts and radiative decay rates for hydrogen-like ions in lowest order perturbation theory. Within the framework of a lowest order non-relativistic treatment of the electron velocities the presented results transcend the dipole approximation. Corresponding results in the dipole approximation can easily be obtained as a limiting case of this theoretical framework (cf. Sec. \ref{Evaldipole}). This algebraic approach is based on a particular representation of the Lie algebra $so(4,2)$ which has been discussed in detail recently by Maclay in connection with  evaluating hydrogenic Lamb shifts in the dipole approximation \cite{Maclay,Maclay1}.

As a main result integral representations have been derived for the complex-valued energy shifts of hydrogenic energy eigenstates with well defined angular momenta. These integral representations transcend the resent results of Maclay \cite{Maclay,Maclay1} as they are not restricted to the dipole approximation and also allow for a unified treatment of Lamb shifts and radiative decay rates.
Within this approach the relevant matrix elements of the unitary effective real-time evolution operator entering these integral representations have been determined algebraically by exploiting characteristic properties of the used $so(4,2)$ representation and by taking advantage of one of its $so(2,1)$ subalgebras and its associated Lie group $SO^+(2,1)=SU(1,1)/{\mathbb Z}_2$. For this purpose advantage has been taken of the fact that the unitary real-time evolution operator is an element of $SO^+(2,1)$ and that
the Lie group $SU(1,1)$ is its double cover. With the help of a well-known representation independent Baker-Campbell-Hausdorff  disentangling relation for $SU(1,1)$ \cite{Chiribella2006,Puri} it has been possible
to determine the required matrix elements of
the unitary real-time evolution operator and to relate them to hypergeometric functions or Jacobi polynomials whose arguments are determined by the coordinate functions of the corresponding two-dimensional representations of $SU(1,1)$ (cf. appendix \ref{efftime1}).

Numerical results for Lamb shifts and radiative decay rates for hydrogen have been presented which demonstrate that the presented integral representations are convenient starting points for evaluating these quantities. In particular, the presented
numerical results for $s$- and $p$-states with principal quantum numbers $N\leq 4$ and their comparison with the corresponding results of the dipole approximation (cf. Sec. \ref{Evaldipole})  demonstrate that the values of the cut-off independent Lamb shifts without dipole approximation tend to lie between the corresponding results of the dipole approximation with and without relativistic corrections (cf. \ref{Evalnum}). The numerical results for the cut-off independent Lamb shifts, which transcend the dipole approximation, are in good agreement with previous results of Seke and M\"odritsch \cite{Seke,Seke1}.
The numerical results for the radiative decay rates in the dipole approximation tend to be slightly larger than the corresponding results without dipole approximation.
Furthermore,
in the dipole approximation
relativistic contributions to Lamb shifts already significantly dominate atomic contributions for $p$-states. These latter atomic contributions are characterized conveniently by Bethe-logarithms, which can also be conveniently evaluated with the help of the presented integral representations for the complex-valued energy shifts. Their numerical values agree well with already known results (cf. Sec. \ref{Evalnum}). The presented integral representations for the complex-valued energy shifts allow for a particularly simple analytical determination of radiative decay rates of hydrogenic energy eigenstates of maximal angular momenta in the dipole approximation (cf. (\ref{Gammacircular})). In particular, they demonstrate that for large values of the principal quantum numbers these rates reduce to previously derived semiclassical results \cite{MarxerSpruch}. These asymptotic results for hydrogenic energy eigenstates of large principal quantum numbers and maximal angular momenta suggest that the presented integral representations of Sec. \ref{LieR} may also be a convenient starting point for deriving asymptotic relations for radiative decay rates and Lamb shifts of energy eigenstates with large principal quantum numbers and small values of angular momenta.

\begin{acknowledgments}
The author thanks A.R.P. Rau for having introduced him to the powerful methods of Lie algebras and for continuous stimulating discussion on dynamical symmetry groups. Furthermore, the author is indebted to Christopher Charnes for illuminating hints on subtleties of group theory and group representations.
\end{acknowledgments}

\appendix
\section{Basic properties of the Lie algebra $so(4,2)$\label{so42}}

For convenience in this appendix characteristic basic properties of the Lie algebra $so(4,2)$ are briefly summarized. They are discussed in detail in numerous references, such as \cite{Maclay,Rau1990,Wybourne,Englefield,Rau}.

A basis of the abstract Lie algebra $so(4,2)$ has $n(n-1)/2$ elements with $n=4+2=6$. They define a totally antisymmetric matrix ${\bf S}$
with components $S_{ij}=-S_{ji}$ $i,j\in \{0,1,2,3,4,5\}$.
The Lie products (commutators) of these elements are defined by
\begin{eqnarray}
\left[S_{ab},S_{cd} \right] &=& i\left(
g_{ac} S_{bd} + g_{bd} S_{ac} - g_{ad} S_{bc} - g_{bc} S_{ad} 
\right)
\label{structure}
\end{eqnarray}
with the corresponding metric tensor ${\bf g}$
being diagonal 
with diagonal matrix elements $g_{11} =g_{22} = g_{33} = g_{44} = 1$ and $g_{55} = g_{00} = -1$.
%
In representations associated with quantum mechanical applications, such as the representation (\ref{so42hydrogen}),  the matrix elements $S_{i j}$ are typically renamed in terms of three triples $\left\{L_i\right\}$, $\left\{A_i\right\}$, $\left\{G_i\right\}$   with $S_{ij}= \epsilon_{ijk}L_k, A_i=S_{i4}, G_i = S_{i0}$ and $i,j,k\in \left\{1,2,3\right\}$, 
a single quantity $G_4=S_{40}$ and a quintuple $\left\{\Gamma_i\right\}$ with $\Gamma_i = S_{i5}$ ($i=0,1,\cdots,4$), i.e.
\begin{eqnarray}
{\bf S} &=&
\left(
\begin{array}{cccccc}
0&-G_1&-G_2&-G_3&-G_4&\Gamma_0\\
G_1&0&L_3&-L_2&A_1&\Gamma_1\\
G_2&-L_3&0&L_1&A_2&\Gamma_2\\
G_3&L_2&-L_1&0&A_3&\Gamma_3\\
G_4&-A_1&-A_2&-A_3&0&\Gamma_4\\
-\Gamma_0&-\Gamma_1&-\Gamma_2&-\Gamma_3&-\Gamma_4&0
\end{array}
\right).
\end{eqnarray}
In general the three operators
\begin{eqnarray}
I_1 &=& -\frac{1}{2}\sum_{i,j=0}^5 S_{ij} S^{ij},~~I_2 = \sum_{i,j,k,l,m,n=0}^5 \epsilon_{ijklmn} S^{ij}S^{kl}S^{mn},~~I_3 = 
\sum_{i,j,k,l=0}^5
S_{ij}S^{jk}S_{kl}S^{li}
\label{Casimir2}
\end{eqnarray}
with $S^{ij} = \sum_{a,b=0}^5g^{ia} g^{jb} S_{ab}$ and $\sum_{b=0}^5g^{ib}g_{bj} = \delta_{ij}$
($i,j \in \{0,\cdots,5\}$)
are a maximal set of independent Casimir operators of this Lie algebra of rank three.
($\epsilon_{abcdef}$
denotes the totally antisymmetric epsilon-symbol). Different values of these three Casimir invariants distinguish different inequivalent representations of this Lie algebra.

Some subalgebras of $so(4,2)$, which are important for our discussion, are:
\begin{itemize}
\item
A small subalgebra of $so(4,2)$ is $so(3)$ with basis elements $\{L_1,L_2,L_3\}$ which fulfil  the characteristic angular momentum commutation relations ($j,k,l\in \{1,2,3\}$)
\begin{eqnarray}
[L_j,L_k] &=& i\epsilon_{jkl}L_l.
\end{eqnarray}
This Lie algebra generates the group $SU(2)$ as a double cover and simultaneously also universal covering group of the identity-connected component of $SO(3)$. 
\item
Another small subalgebra of $so(4,2)$ is $so(2,1)\simeq su(1,1)$ with basis elements $\{J_1=\Gamma_4, J_2 = G_4, J_3 = \Gamma_0 \}$ with the commutation relations ($a,b,c\in \{1,2,3\}$)
\begin{eqnarray}
[J_a,J_b] &=& i\epsilon_{abc}J_c \gamma_{cc}
\end{eqnarray}
with $\gamma_{33}= -1, \gamma_{22} = \gamma_{11} = 1$. 
This Lie algebra generates the group $SU(1,1)$, for example, which is a double cover (but not universal covering group) of the identity-connected component of $SO(2,1)$.
\end{itemize}

\section{Derivation of the sum rule of (\ref{sumrule}) and of the mass renormalization term of (\ref{Delta1})\label{sumrule1}}
In this appendix the derivation of the sum rule (\ref{sumrule}), which has been discussed recently by Maclay \cite{Maclay}, is summarized and generalized to arbitrary representations of the Lie algebra $so(4,2)$. It is also demonstrated that the quantity $\Delta$ of (\ref{Delta1}) only contributes to the mass renormalization.

Following the notation of  appendix \ref{so42} and Sec. \ref{LieH} let us consider a quintuple $n:= (n_0,n_1,n_2,n_3,n_4)^T \in {\mathbb R}^5$ and define
$n^T\Gamma (a)= \sum_{i=0}^4\Gamma_i (a) n_i$ with $a>0$. Furthermore, for $n^T g n = \sum _{i=0}^4 g_{ii }n_i^2 $ 
 with $1=-g_{00}=g_{11}=g_{22}=g_{33}=g_{44}$ let us introduce the parameter dependent operators
\begin{eqnarray}
\Gamma_i (T,a) &=&
e^{in^T \Gamma (a) T} \Gamma_i(a)  e^{-in^T \Gamma (a) T}, ~S_{ij}(T,a) = 
e^{in^T \Gamma (a) T} S_{ij}(a) e^{-in^T \Gamma (a) T},~~i,j\in \{0,1,\cdots,4\}.
\end{eqnarray}
From the commutation relations of $so(4,2)$ we find
\begin{eqnarray}
\Gamma_i (T,a) &=& \Gamma_i (a)\cosh \left(T\sqrt{-g_{55} n^T g n}\right)  -
\frac{g_{55}\sum_{j=0}^{4}( n_j S_{ji}(a))}{\sqrt{-g_{55}n^T g n}}
 \sinh\left(T\sqrt{-g_{55} n^T g n}\right) -\nonumber\\
 &&
 \frac{g_{55}g_{ii} n_i (n^T\Gamma(a))}{(-g_{55} n^T g n)}\left(1-
\cosh \left(T\sqrt{-g_{55} n^T g n}\right) 
  \right)
\end{eqnarray}
by solving the differential equations
\begin{eqnarray}
\frac{\partial ^2}{\partial T^2} \Gamma_i(T,a) &=& -g_{55}\left((n^T g n ) \Gamma_i(T,a) - (n^T \Gamma(a)) g_{ii} n_i\right),\nonumber\\
\frac{\partial \Gamma_i}{\partial T}(T,a)&=&-g_{55}\sum_{j=0}^4 n_j S_{ji}(T,a).
\end{eqnarray}
 This differential equation is a consequence of the vector character of the $5$-vector $\Gamma(a)$ under $so(4,2)$ and the commutation relations $[\Gamma_k(a),\Gamma_j(a)]= ig_{55} S_{kj}(a)$.
From these latter commutation relations we find
$\sum_{i=0}^4 g_{ii}(\Gamma_i(a) S_{ij}(a) + S_{ij}(a) \Gamma_i(a)) =0$ and from the vector character of $\Gamma (a)$ we find $\sum_{i=0}^4 g_{ii} (\Gamma_i(a) S_{ij}(a) -S_{ij}(a) \Gamma_i(a)) = -4i\Gamma_j(a)$, which finally leads to the relation
\begin{eqnarray}
\sum_{i=0}^4 g_{ii}S_{ij}(a) \Gamma_i (a)&=&
-\sum_{i=0}^4 g_{ii}\Gamma_i(a) S_{ij}(a) =
2i \Gamma_j(a).
\end{eqnarray}
Using this latter relation we find the general sum rule
\begin{eqnarray}
&&\sum_{i=0}^4 g_{ii} \Gamma_i(a)
e^{-in^T \Gamma (a) T}
\Gamma_i(a) =
\sum_{i=0}^4 g_{ii}
e^{-in^T \Gamma(a) T}
\Gamma_i (T,a) \Gamma_i(a) =\\
&&
e^{-in^T \Gamma (a) T}
\left(
\left(\Gamma^T(a) g \Gamma(a) -\frac{(n^T\Gamma(a))^2}{n^T g n}\right) \cosh \left(T\sqrt{-g_{55} n^T g n}\right) +
\frac{2i g_{55} (n^T\Gamma(a))}{\sqrt{-g_{55} n^T g n}}\sinh \left(T\sqrt{-g_{55} n^T g n}\right)
+ \frac{(n^T\Gamma(a))^2}{n^T g n}
\right)\nonumber
\end{eqnarray}
with $\Gamma^T(a) g \Gamma (a)= \sum_{i=0}^4 (\Gamma_i(a))^2 g_{ii}$.
If $-g_{55} n^T g n < 0$  with
$|g_{55}| = 1$
this sum rule reduces to 
\begin{eqnarray}
&&\sum_{i=0}^4 g_{ii} \Gamma_i(a)
e^{-in^T \Gamma (a) T}
\Gamma_i(a) =
e^{-i(n^T \Gamma(a) -\sqrt{|n^T g n}|)T}
\frac{1}{2}\left(
\Gamma^T(a)g \Gamma(a)  + g_{55} - g_{55}
\left(
\frac{n^T \Gamma(a)}{\sqrt{|n^T g n|}}-1
\right)	^2
 \right) +\nonumber\\
 &&
 e^{-i(n^T \Gamma(a) +\sqrt{|n^T g n}|)T}
\frac{1}{2}\left(
\Gamma^T(a) g \Gamma(a)  + g_{55} - g_{55}
\left(
\frac{n^T \Gamma(a)}{\sqrt{|n^T g n|}}+1
\right)	^2
 \right)
 +e^{-in^T \Gamma(a) T}\frac{(n^T\Gamma(a))^2}{n^T g n} =\nonumber\\
&& \left(\Gamma^T(a) g \Gamma(a) + g_{55} \right) e^{-in^T \Gamma(a) T}\cos(\sqrt{|n^Tgn|}T) -2
\frac{g_{55}}{|n^Tg n|}\frac{d^2}{dT^2}\left(\sin^2(\sqrt{|n^T g n|}T/2) e^{-in^T \Gamma(a) T}\right).
\end{eqnarray}
According to the particular representation of $so(4,2)$ (cf. (\ref{so42hydrogen})) we have the constraint $\Gamma^2 (a)=\Gamma^T(a) g \Gamma(a) =1$ so that in this case we obtain the simplified sum-rule 
\begin{eqnarray}
&&\sum_{i=0}^4 g_{ii}\Gamma_i(a) (n^T\Gamma (a) - \nu - i0)^{-1} \Gamma_i(a) = i\sum_{i=0}^4 g_{ii}\Gamma_i(a) \int_0^{\infty} dT e^{-in^T\Gamma (a) T}e^{i(\nu+i0) T}
\Gamma_i(a) 
=\nonumber\\
&&
 i\int_0^{\infty}dT e^{i(\nu+i0) T}\frac{(-2g_{55})}{|n^T g n|}
 \frac{d^2}{dT^2}\left(\sin^2(\sqrt{|n^T g n|}T/2)
e^{-in^T\Gamma (a) T}\right)\nonumber\\
\label{shift-general}
\end{eqnarray}
because $g_{55}=-1$, and in the special case $|n^T g n|=1$ one obtains relation (\ref{sumrule}).

Let us consider the quantity
\begin{eqnarray}
\Delta &=&
\lim_{\epsilon \to 0}\frac{m_e\nu}{N^2}
(NLM;a_N|\left(\Gamma_0(a_N) \left(n^T\Gamma(a_N) - \nu -i\epsilon\right)^{-1}\Gamma_0(a_N) - \Gamma_4(a_N) \left(n^T\Gamma(a_N) - \nu -i\epsilon \right)^{-1}\Gamma_4(a_N)\right)|NLM;a_N)\nonumber\\
\end{eqnarray}
with $\nu = Ne^{-\Phi}$
and $n =(\cosh\Phi,0,0,0,-\sinh \Phi)^T$. From (\ref{scaling1}) we obtain the relation
\begin{eqnarray}
\left(\Gamma_0(a_Ne^{\Phi}) - \nu\right) |NLM;a_N) &=& \left( n^T \Gamma(a_N) - \nu\right)|NLM;a_N) = \sinh \Phi \left(N-\Gamma_4(a_N)\right) |NLM;a_N)
\end{eqnarray}
which implies
\begin{eqnarray}
\Delta &=& \frac{m_e\nu}{N^2\sinh^2 \Phi}(NLM;a_N|\left(n^T\Gamma(a_N)-\nu\right)|NLM;a_N)
\end{eqnarray}
for $\Phi \neq 0$. Using $(NLM;a_N|\Gamma_4(a_N) |NLM;a_N) = 0$, (\ref{eigenvalue1}) and $E = E_Ne^{2\Phi}$ with $E_N= - (\hbar a_N)^2/(2m_e) = -(\hbar a_1)^2/(2m_eN^2)$ ($\hbar a_1=m_ecZ\alpha_0 $) one obtains the result
\begin{eqnarray}
\Delta &=& \frac{2m_eE_N}{E-E_N}.
\end{eqnarray}
The virial theorem states $\langle NLM| \left({\bf P}^2/m_e\right)|NLM\rangle = \langle NLM|\left( Ze^2/(4\pi\epsilon_0r)\right)|NLM\rangle = - 2E_N$ (\cite{CohenTannoudji1}) so that we finally obtain the result
\begin{eqnarray}
\Delta &=& \frac{\langle NLM| {\bf P}^2|NLM\rangle}{E_N-E}
\end{eqnarray}
for $E_N\neq E$. Thus $\Delta$ contributes to the mass renormalization.

\section{Derivation of the matrix elements (\ref{hyper}) of the effective-time evolution operator \label{efftime1}}
In this appendix relation (\ref{hyper}), a main results of this paper, is derived with the help of group theoretical properties of the Lie group $SU(1,1)$, a double cover of the identity-connected component $SO^+(2,1)$ of $SO(2,1)$. The Lie algebra $su(1,1)\simeq so(2,1)$ of these groups is a subalgebra of $so(4,2)$. For the sake of convenience and for fixing notation and arbitrary phase conventions in a consistent way we first summarize basic arguments leading to the classification of all irreducible unitary representations of the Lie group $SU(1,1)$ with lower-bounded spectra of its generators. In a second step we use these properties to evaluate the matrix elements of the effective time evolution operator entering (\ref{complexshiftfinal}) and (\ref{shiftnumerical}) with the help of a Baker-Campbell-Hausdorff relation for the Lie group $SU(1,1)$. 

\subsection{Irreducible unitary representations of $SU(1,1)$\label{grouptheory}}
Let us start with a definition of the continuous Lie group $SU(1,1)$ whose defining representation consists of all complex valued
 $2\times 2$ matrices of the form
\begin{eqnarray}
u(\alpha,\beta)&=&
\left(
\begin{array}{cc}
\alpha & \beta\\
-g_{33}\beta^*&\alpha^*
\end{array}\right)
~~\longrightarrow~~
u^{-1}(\alpha,\beta) =
\left(
\begin{array}{cc}
\alpha^* & -\beta\\
g_{33}\beta^*&\alpha
\end{array}\right)
\label{group}
\end{eqnarray}
with ${\rm Det}\left(u(\alpha,\beta)\right) = ~\mid\alpha\mid^2 - \mid\beta\mid^2= 1$; $\alpha,\beta\in {\mathbb C}$ and $g_{33} = -1$. The corresponding matrices with $g_{33} = 1$ would constitute the defining representation of the Lie group $SU(2)$. Introducing real valued coordinates $(t,x,y,z)$ by the linear relations $\alpha=t+iz,\beta =x+iy$ with $(t,z,x,y)^T\in {\mathbb R}^4$ the elements of $SU(1,1)$ and $SU(2)$ are characterized by the corresponding constraints ${\rm Det}\left(u(\alpha,\beta)\right) = t^2 +z^2 + g_{33}(x^2+y^2) =1$. 

Complete linear independent sets of generators of this matrix group are given by the $2\times 2$ matrices
\begin{eqnarray}
\frac{du}{dz}\mid_{(x,y,z)=(0,0,0)} &=&i\sigma_3,~
\frac{du}{dx}
\mid_{(x,y,z)=(0,0,0)}
 =
\sigma_1
,~
\frac{du}{dy}
\mid_{(x,y,z)=(0,0,0)}
 = 
-\sigma_2
\end{eqnarray}
with $\sigma_i,~i\in \left\{ 1,2,3\right\}$ denoting the Pauli spin matrices \cite{CohenTannoudji1}.
Thus, the Lie algebra of the $2\times 2$ matrices 
$
j_3 = \sigma_3/2, j_1 = i\sigma_1/2,
j_2 = i\sigma_2/2
$ fulfils the characteristic commutation relations (\ref{comm}) or equivalently in terms of the ladder operators $j_{\pm} = i\sqrt{1/2}(j_1\pm i j_2)$ the relations
\begin{eqnarray}
\left[j_+ ,j_- \right] &=& j_3 ,~
\left[j_3 ,j_{\pm}\right] =  \pm j_{\pm}.
\label{ladder}
\end{eqnarray}
The operator
\begin{eqnarray}
j^2 &=& j_1^2 + j_2^2-j_3^2 = -\left(j_3^2 \pm j_3 + 2 j_{\mp}j_{\pm}
 \right).
 \label{Casimir3}
\end{eqnarray}
 is the only independent Casimir operator of this semi-simple Lie algebra of rank 1.
 This two-dimensional matrix representation of the abstract Lie algebra $su(1,1)$ is not unitary reflecting the fact that the associated Lie group $SU(1,1)$ is non compact, as a unitary representation of a non-compact Lie group necessarily has to be infinite dimensional. 
 
 In a unitary infinite dimensional representation of the Lie group $SU(1,1)$ with respect to a given scalar product the generators $J_k$ ($k=1,2,3$) of unitary transformations of the form $U(\lambda)= e^{i\lambda J_k}$, corresponding to the two-dimensional generators $j_1.j_2,j_3$, have to be hermitian, i.e.
$J_1 = J_1^{\dagger}, 
J_2 = J_2^{\dagger},
J_3 = J_3^{\dagger}, J^{2\dagger} = J^2$, which implies the relations
\begin{eqnarray}
J_{\pm}^{\dagger} = g_{33}J_{\mp}
\label{unitary}
\end{eqnarray}
for the ladder operators.
A representation space  of a unitary representation, say ${\cal D}^{(X)}$, with an orthonormal basis, say $\left\{ |X m)\right\}$, can be constructed by diagonalizing the complete set of commuting operators $J^2, J_3$ simultaneously according to (\ref{simultaneous}).
Relations (\ref{ladder}) are valid for the corresponding hermitian generators $J_1,J_2,J_3$ and imply
\begin{eqnarray}
J_3 J_{\pm} |X~m) &=& \left( m \pm 1\right) J_{\pm}|X~m).
\end{eqnarray}
Together with the unitarity condition (\ref{unitary}) they yield the characteristic constraints
\begin{eqnarray}
0 \leq 2||J_{\pm}|X~m)||^2
 &=&\left(X-g_{33}m(m\pm 1) \right) 
 |||X~m)||^2,
 \label{norm}
 \\
||J_{+}|X~m)||^2 &=&-g_{33}m
|||X~m)||^2
+
||J_{-}|X~m)||^2
\label{sign}
\end{eqnarray}
with the scalar product-induced norm
$|||X~m)|| = 
\sqrt{(X~m|X~m)}$.
In addition, relations (\ref{ladder}) imply
\begin{eqnarray}
( X~m'|\left[J_3 , J_{\pm}\right]|X~m) &=&
\pm ( X~m'|J_{\pm}|X~m) =
(m'-m)
( X~m'|J_{\pm}|X~m),
\end{eqnarray}
so that $|m'-m| =1$ for states with 
$( X~m'|J_{\pm}|X~m)\neq0$.
Thus, according to (\ref{norm}) all normalized states with $( X~m|X~m) =1$ of a unitary representation of $SU(1,1)$ have to fulfil the relation
\begin{eqnarray}
2||J_{\pm}|X~m\rangle||^2 &=& X+m(m\pm 1) \geq 0.
\label{condition}
\end{eqnarray}
For a given value of $X$ relation (\ref{condition}) can always be fulfilled for a sufficiently large positive 
value of $m$, say $m_1$, or for a sufficiently small negative value of $m$. Thus, let us assume that there exists a corresponding normalized state $|X~m_1\rangle$.
Applying the ladder operator $J_-$ successively to the state $|X~m_1)$ with $m_1 > 0$ produces  the states $\left\{|X~m_1 -k)| k\in {\mathbb N}_0  \right\}$. Analogously, if $m_1 $ were a sufficiently small negative value so that (\ref{condition}) is fulfilled,
applying the ladder operator $J_+$ successively to the state $|X~m_1)$ would produce  the states $\left\{|X~m_1 +k ) | k\in {\mathbb N}_0  \right\}$.
Therefore, at least two different cases of unitary representation spaces can be distinguished, namely discrete spectra bounded from below and discrete spectra bounded from above.

For our discussion discrete spectra bounded from below are particularly important. Thus, we restrict our subsequent discussion to this case. Correspondingly, let us consider a sequence of states 
$\left\{|X~m_1 -k)| k\in {\mathbb N}_0  \right\}$ 
which terminates. Therefore, there is a normalized state, say $|X~m_0)$, with $J_-|X~m_0) =0$ so that (\ref{norm}) implies 
\begin{eqnarray}
X&=& m_0 (1-m_0).
\label{X}
\end{eqnarray}
If one is interested in irreducible unitary representations, relation (\ref{sign}) imposes the additional condition  $m_0 \geq 0$. This is due to the fact that in an irreducible representation each state is cyclic, i.e.
all states of the representation can be reached from an arbitrarily given state by successive applications of the operators of the Lie algebra. 
If $m_0$ were negative, the normalized state $|X m_0)$ could be reached from an arbitrary normalized state of the representation by successive applications of the ladder operators $J_-$, but according to (\ref{sign}) a normalized state $J_+|X m_0)$ would not exist. Therefore, it is impossible to reach the other normalized states of the representation by successive application of the ladder operator $J_+$ to the normalized state $|X m_0)$, and  $m_0 < 0$ yields indecomposable unitary representations.
Thus, the lower bounded spectra of $J_3$ of all possible unitary irreducible representations are necessarily of the form
\begin{eqnarray}Sp(J_3) = \left\{m_0+k | k \in {\mathbb N}_0, 
0 \leq m_0 \in {\mathbb R}
 \right\},
 \label{spektrum}
\end{eqnarray}
 because the states produced by successive applications of the ladder operator $J_+$ to the state $|X~m_0)$ do not terminate. 
 The special case $m_0=0$ characterizes the trivial identity representation for which  $J_+|X~m_0)= J_-|X~m_0) =0$.

 All the conclusions leading to (\ref{spektrum}) can be drawn from basic properties of the Lie algebra $su(1,1)$ alone.
However, as the associated Lie group $SU(1,1)$ is not simply connected
further restrictions on the possible values of $m_0$ arise from global considerations of this group \cite{Biedenharn}. 
For this purpose let us consider the compact maximal Abelian $U(1)$ subgroup  with elements $u_1(\varphi) = e^{i\varphi \sigma_3/2}\in U(1)\subset SU(1,1)$ with $0\leq \varphi \leq 4\pi$ and the global property $u_1(4\pi) = u_1(0)$. With respect to this $U(1)$ subgroup every irreducible unitary representation of $SU(1,1)$ decomposes into a direct sum of one-dimensional irreducible unitary representations of this $U(1)$ subgroup. It is known that every irreducible representation of this maximal compact subgroup $U(1) \subset SU(1,1)$ occurs at most once in every irreducible unitary representation of $SU(1,1)$ \cite{Godement}. Thus, the global property
$u_1(4\pi) = u_1(0)$ of this compact maximal Abelian subgroup $U(1)$ 
 imposes the further constraint 
 \begin{eqnarray}
 0 \leq 2m_0 \in {\mathbb Z}
 \label{globalconstraint}
 \end{eqnarray}
 for possible irreducible unitary representations of $SU(1,1)$. Contrary to the case of $SU(2)$ this quantization condition does not follow already from the properties of the Lie algebra $su(1,1)$, i.e. from the local properties of $SU(1,1)$. This is due to the fact that $SU(1,1)$ is not simply connected and is thus not its own universal covering group. As universal covering groups of connected groups are always simply connected, the properties of their Lie algebras already determine their global properties uniquely. The group manifold of the universal covering group of $SU(1,1)$ is ${\mathbb R}_2\times {\mathbb R}$ covering the group manifold of $SU(1,1)$ infinitely many times.
 
In summary, the possible discrete spectra of $J_3$ of
irreducible unitary representations of $SU(1,1)$, which are bounded from below, are given by (\ref{spektrum}) with (\ref{X}) and with the additional global constraint (\ref{globalconstraint}). For integer values of $m_0$ the resulting irreducible unitary representations of $SU(1,1)$ are irreducible unitary representations of $SO^+(2,1)=SU(1,1)/{\mathbb Z}_2$, whereas for half-integer values of $m_0$ they are projective representations of $SO^+(2,1)$.
Thus, with the identifications $J_3\to \Gamma(a_N)$ and  $m_0 \to L+1$  for each fixed value of the angular momentum quantum numbers $L\in {\mathbb N}_0$ and $M\in {\mathbb Z}$ with $-L \leq M\leq L$ the orthonormal states 
$\{|n L M;a_N)|~ L+1\leq n\in {\mathbb N}\}$ of (\ref{eigenGamma0}) constitute an irreducible unitary representation space of the group $SO^+(2,1)=SU(1,1)/{\mathbb Z}_2$ which is the identity-connected component of $SO(2,1)$.

 \subsection{Group theoretical evaluation of the matrix elements (\ref{hyper}) and  (\ref{Rn})}
In this appendix it is demonstrated that basic representation independent group properties of $SU(1,1)$ enable an evaluation of 
the matrix elements of the effective time evolution operator entering (\ref{Q}). For this purpose advantage is taken
of a well known Baker-Campbell-Hausdorff relation for the group $SU(1,1)$ which is also frequently used in different physical contexts, such as quantum state tomography  \cite{Chiribella2006,Puri}.

With the help of an orthonormal basis of an irreducible unitary representation space ${\cal D}^{(X)}$ all possible unitary transformations $U(u(\alpha,\beta))$ with $u(\alpha,\beta)\in SU(1,1)$ can be constructed by exponentiation of linear combinations of the three linear independent hermitian generators $J_1,J_2,J_3$ which correspond to the $2\times 2$ matrices $j_1= i\sigma_1/2, j_2 = i\sigma_2/2,j_3 = \sigma_3/2$ of the non-unitary matrix representation (\ref{group}).
For the evaluation of matrix elements of $U(u(\alpha,\beta))$ the relations
 \begin{eqnarray}
 J_{\pm} |X~m) &=& \pm \sqrt{\frac{1}{2}}\sqrt{m_0 (1-m_0) + m (m \pm 1)}|X~m\pm 1) =
\pm \sqrt{\frac{1}{2}}\sqrt{(m_0 \pm m)(1\pm m-m_0)}|X~m\pm 1)
 \label{matrixelement}
 \end{eqnarray}
 can be obtained from (\ref{norm}).
They are valid for the orthonormal states $|X~m)$ and $ |X~m\pm 1)$ of the irreducible unitary representation ${\cal D}^{(X)}$ characterized by the eigenvalue $X$ of (\ref{X}). Thereby, the arbitrary phase of these states has been chosen in such a way that 
 $( X~m+1| J_+ |Xm ) > 0$. Note that consistent with this phase convention  the sign change in (\ref{matrixelement}) is required by the hermiticity requirement (\ref{unitary}).
Let us now use the representation independent Baker-Campbell-Hausdorff relation \cite{Chiribella2006,Puri}
\begin{eqnarray}
U(u(\alpha,\beta)) &=& 
e^{-\left(J_+\sqrt{2}e^{i(\chi +\psi)} \tanh\rho\right)}
\left(\frac{1}{\cosh \rho}\right)^{2J_3}
e^{-\left(J_-\sqrt{2} e^{-i(\chi+\psi)}\tanh \rho\right)}e^{2i\chi J_3}
\end{eqnarray}
with 
\begin{eqnarray}
\alpha &=& e^{i\chi}\cosh \rho,~~\beta = e^{i\psi}\sinh \rho ,
\end{eqnarray}
$\rho  \geq 0$ and $ \chi,\psi \in \left[0,2\pi \right.)$. As the validity of this Baker-Campbell-Hausdorff relation only relies on the properties of the Lie algebra $su(1,1)$, it is representation independent and can thus be easily proved by using the defining non-unitary representation (\ref{group}) of $SU(1,1)$, for example.
As a result matrix elements of $U(u(\alpha,\beta))$ with respect to the orthonormal basis states $\left\{|X~m)\right\}$ of ${\cal D}^{(X)}$  are given by ($m,m'\geq m_0$)
\begin{eqnarray}
( X~m'| U(u(\alpha,\beta)) | X~m ) &=&
\sum_{r=m_0}^{\min(m,m')}
 ( X~m'|
e^{-\left(J_+\sqrt{2}e^{i(\chi +\psi)} \tanh\rho\right)}
|X~r) \frac{e^{2i\chi m}}{(\cosh \rho)^{2r}}( X~r|
e^{-\left(J_-\sqrt{2} e^{-i(\chi+\psi)}\tanh \rho\right)}|X~m)
 =\nonumber\\
&&
\sum_{r=m_0}^{\min(m,m')}
\frac{\left(-e^{i(\chi + \psi)}\tanh\rho\right)^{m'-r}}{(m'-r)!}
( X~m'| \left(\sqrt{2}J_+ \right)^{m'-r}|X~r)
\frac{e^{2i\chi m}}{(\cosh \rho)^{2r}}\times\nonumber\\
&&\hspace{1cm}
\frac{\left(-e^{-i(\chi+\psi)}\tanh \rho
\right)^{m-r}}{(m-r)!}( X~r| \left(\sqrt{2}J_- \right)^{m-r}|X~m) =\nonumber\\
&&
\sum_{r=m_0}^{\min(m,m')}
\left(-e^{i(\chi + \psi)}\tanh\rho
\right)^{m'-r}
\left(
e^{-i(\chi + \psi)}\tanh\rho
\right)^{m-r}
\frac{e^{2i\chi m}}{(\cosh \rho)^{2r}}\times\nonumber\\
&&
\frac{\sqrt{\Gamma(m'+m_0)\Gamma(m'-m_0+1)\Gamma(m-m_0+1)\Gamma(m+m_0)}}{\Gamma(r+m_0)\Gamma(r-m_0+1)\Gamma(m'-r+1)\Gamma(m-r+1)}
\label{representation}
\end{eqnarray}
with $\Gamma(m+1)=m!$ \cite{AS}.
For $m'\leq m$
this relation yields the matrix elements of the representation ${\cal D}^{(X)}$ in the form
\begin{eqnarray}
( X~m'| U(u(\alpha,\beta)) | X~m) &=&\frac{(\beta^*)^{m-m'}}{(\alpha^*)^{m+m'}}\sqrt{\frac{\Gamma(m+m_0)\Gamma(m-m_0+1)}{\Gamma(m'+m_0) \Gamma(m'-m_0+1)}}\times\nonumber\\
&&
\frac{{_2 F_1}\left(m_0-m',-m_0+1 -m';m-m'+1;-|\beta|^2\right)}{\Gamma(m-m'+1)}
\label{finitetrafo}
\end{eqnarray}
with the hypergeometric function ($|z|<1$)
\begin{eqnarray}
{_2 F_1}(a,b;c;z) &=& \frac{\Gamma(c)}{\Gamma(a)
\Gamma(b)}\sum_{n=0}^{\infty}\frac{\Gamma(a+n)\Gamma(b+n)}{\Gamma(c+n)}\frac{z^n}{n!}
\end{eqnarray}
and with $|\beta|^2 = |\alpha|^2-1$.
Thereby, we have used the reflection formula for the Gamma function \cite{AS}, i.e.
\begin{eqnarray}
\frac{\Gamma(n-A)}{\Gamma(-A)} &=& (-1)^n \frac{\Gamma(1+A)}{\Gamma(1+A-n)}.
\end{eqnarray}
Apart from a direct evaluation, matrix elements with $m \leq m'$ can more conveniently be evaluated from (\ref{finitetrafo}) with the help of the unitarity condition of the representation ${\cal D}^{(X)}$, i.e.
\begin{eqnarray}
( X~m'| U(u^{-1}(\alpha,\beta))|X~m)^* &=&
( X~m| U(u(\alpha,\beta))|X~m'), ~~{\rm i.e.}\nonumber\\
U(u^{-1}(\alpha,\beta)) &=& U^{\dagger}(u(\alpha,\beta))
\end{eqnarray}
with $u^{-1}(\alpha,\beta) = u(\alpha^*,-\beta)$ (cf.(\ref{group})). As  (\ref{representation}) is invariant under the transformation $m_0 \to \tilde{m}_0 = 1-m_0$ with $X=m_0(1-m_0) = \tilde{m}_0(1-\tilde{m}_0)$, $m_0$ and $\tilde{m}_0$ yield equivalent unitary representations.

With the identifications
$\Gamma_4(a_N) := J_1, G_4 := J_2, \Gamma_0(a_N) := J_3$ the three operators
$\left\{\Gamma_4(a_N), G_4, \Gamma_0(a_N)\right\}$ constitute a particular representation of the subalgebra $su(1,1)\subset so(4,2)$ with Casimir operator $J^2 = -{\bf L}^2$. Thus, from the eigenvalue relation (\ref{eigenvalue1}) and the discussion of appendix \ref{grouptheory}
it is apparent that for
fixed values of $L\in {\mathbb N}_0$ and $M\in {\mathbb Z}$ with $-L\leq M\leq L$ 
the orthonormal states 
$\left\{|nLM;a_N)\right\}$ of (\ref{eigenGamma0}) 
constitute an irreducible unitary representation of $SO^+(2,1)=SU(1,1)/{\mathbb Z}_2$
with $m_0 = L+1$ and $L+1\leq n\in {\mathbb N}$.

For an evaluation of the matrix element $(NLM;a_N |e^{-i\Gamma_0(a_N e^{\Phi})T}|NLM;a_N)$ one can use (\ref{finitetrafo}) if the coordinate functions, say $(\alpha_0(T,\Phi),\beta_0(T,\Phi))$, are known which describe the unitary effective time evolution operator $e^{-i\Gamma_0(a_N e^{\Phi})T}$. For this purpose we use (\ref{scaling1}) with $\lambda = \Phi$, i.e.
\begin{eqnarray}
\Gamma_0(a_Ne^{\Phi}) &=& e^{i\Phi G_4}\Gamma_0(a_N) e^{-i\Phi G_4} = \Gamma_0(a_N) \cosh \phi - \Gamma_4(a_N)\sinh \Phi.
\label{scaling3}
\end{eqnarray}
This implies
\begin{eqnarray}
e^{-iT\Gamma_0(a_N e^{\Phi})} &=& U(u(\alpha_0(T,\Phi),\beta_0(T,\Phi)))
\end{eqnarray}
with
\begin{eqnarray}
u(\alpha_0(T,\Phi),\beta_0(T,\Phi)) &=& e^{-iT\left(j_3\cosh\phi -j_1 \sinh\Phi \right)} =
\left(
\begin{array}{cc}
\cos(T/2)-i\sin(T/2) \cosh\Phi& -\sin(T/2)\sinh\Phi\\
-\sin(T/2)\sinh\Phi&
\cos(T/2)+i\sin(T/2) \cosh\Phi\\
\end{array}
\right)\nonumber
\end{eqnarray}
with $j_3=\sigma_3/2, j_1= i\sigma_1/2$. Therefore, we obtain the coordinate functions 
\begin{eqnarray}
\alpha_0(T,\Phi) = \cos(T/2)-i\sin(T/2) \cosh\Phi, ~\beta_0(T,\Phi) =-\sin(T/2)\sinh\Phi. \end{eqnarray}
Inserting these coordinate functions into (\ref{finitetrafo}) we finally arrive at (\ref{hyper}).

Analogously one obtains the result
\begin{eqnarray}
e^{i\Phi G_4}&=& U(u(\alpha_1(\Phi),\beta_1(\Phi)))
\end{eqnarray}
with
\begin{eqnarray}
u(\alpha_1(\Phi),\beta_1(\Phi)) &=& 
e^{i \Phi j_2} =
\left(
\begin{array}{cc}
\cosh(\Phi/2)& i\sinh(\Phi/2)\\
-i\sinh(\Phi/2)&
 \cosh(\Phi/2)\\
\end{array}
\right)\nonumber
\label{scaling2}
\end{eqnarray}
and with the coordinate functions
\begin{eqnarray}
\alpha_1(\Phi) &=& \cosh (\Phi/2),~\beta_1(\Phi) = i\sinh(\Phi/2)\end{eqnarray}
from which all matrix elements of a particular representation can be obtained from (\ref{finitetrafo}).
Therefore, using (\ref{scaling3}) one arrives at an alternative expression for (\ref{hyper}), namely
\begin{eqnarray}
(NLM;a_N|e^{-iT\Gamma_0(a_N e^{\Phi})}|NLM;a_N) &=&\sum_{n=L+1}^{\infty} 
\mid (NLM;a_N|e^{i\Phi G_4}|nLM;a_N)\mid ^2 e^{-inT},
\end{eqnarray}
which involves the spectral representation of $e^{-iT\Gamma_0(a_N)}$ and from which (\ref{Rn}) can be obtained.

\end{document}